\documentclass[a4paper,11pt]{article}
\usepackage{pos}

\title{Axions as Dark Matter, Dark Energy, and Dark Radiation}

\author*{Luca Visinelli}
\affiliation{Dipartimento di Fisica ``E.R.\ Caianiello'', Universit\`a degli Studi di Salerno,\\ Via Giovanni Paolo II, 132 - 84084 Fisciano (SA), Italy}
\affiliation{Istituto Nazionale di Fisica Nucleare - Gruppo Collegato di Salerno - Sezione di Napoli,\\ Via Giovanni Paolo II, 132 - 84084 Fisciano (SA), Italy}
\emailAdd{lvisinelli@unisa.it}

\abstract{Axions and axion-like particles are ubiquitous in extensions of the Standard Model and offer a unifying framework for addressing open problems in cosmology. Depending on their mass and interactions, axions can act as dark matter, drive cosmic acceleration as dark energy, or contribute to the relativistic background as dark radiation. Motivated by the plenary talk at TAUP 2025, this proceeding reviews the phenomenology of light bosons in the early and late Universe, with a focus on the theoretical foundations, observational signatures, and experimental prospects. This contribution is intended as a compact mini-review, emphasizing representative mechanisms and observational targets rather than an exhaustive survey.}

\FullConference{19th International Conference on Topics in Astroparticle and Underground Physics (TAUP2025)\\
24–30 Aug 2025\\
Xichang, China\\}

\begin{document}
\maketitle

\section{Introduction}

One of the central achievements of modern cosmology is the establishment of the standard $\Lambda$CDM model, which describes a Universe composed of ordinary baryonic matter, cold dark matter (DM)~\cite{Bertone:2004pz, Marsh:2024ury, Cirelli:2024ssz}, and a cosmological constant ($\Lambda$) driving accelerated expansion~\cite{SDSS:2005xqv, Planck:2018vyg}. Despite its success in accounting for a wide range of observations~\cite{Planck:2019nip}, the $\Lambda$CDM framework leaves fundamental questions unanswered. What is the identity of dark matter? Why is today’s Universe dominated by a mysterious dark energy component? Could additional relativistic species exist beyond the three active neutrinos? Such puzzles motivate the search for new physics beyond the Standard Model (SM) of particle physics. Among the most compelling candidates are \textbf{axions} and their generalizations, axion-like particles (ALPs). The QCD axion was originally proposed by Peccei and Quinn as a dynamical solution to the strong CP problem in QCD~\cite{Peccei:1977hh}. Their mechanism promotes the CP-violating $\theta$ parameter to a dynamical field that relaxes to a CP-conserving minimum. The associated pseudo–Nambu–Goldstone boson was independently identified by Weinberg and Wilczek and named the axion~\cite{Weinberg:1977ma,Wilczek:1977pj}. While motivated by particle physics, axions soon acquired a central role in cosmology: their non-thermal production via the misalignment mechanism can account for the observed DM abundance~\cite{Preskill:1982cy,Abbott:1982af,Dine:1982ah}. Beyond misalignment, axion cosmology is enriched by the dynamics of Peccei–Quinn topological defects. If the $U(1)_{\rm PQ}$ symmetry is broken after inflation, causally disconnected regions select independent initial phases, leading via the Kibble mechanism to the formation of a cosmic string network~\cite{Kibble:1976sj,Vilenkin:1982ks,Lyth:1992tw}. As the Universe expands, this network evolves under Hubble damping and intercommutation, radiating energy predominantly into axions~\cite{Davis:1986xc,Harari:1987ht,Davis:1989nj,Battye:1994au}. Around the QCD epoch, when the axion potential turns on, a discrete set of vacua appears, generating domain walls bounded by strings. The eventual annihilation of this string–wall system releases the bulk of its energy into a stochastic population of cold axions~\cite{Chang:1998tb, Hiramatsu:2010yn,Hiramatsu:2012gg, Kawasaki:2018bzv,Saikawa:2012uk,Vaquero:2018tib}. In some models, domain wall decay can even dominate axion production~\cite{Kawasaki:2014sqa,Ringwald:2015dsf,Harigaya:2018ooc,Caputo:2019wsd, Beyer:2022ywc}. Taken together, the misalignment mechanism and topological defect decays define the standard post-inflationary cosmological history of field-theoretic QCD axions. More broadly, in ALP extensions where the axion mass and couplings are not fixed by the QCD relation, a wide variety of cosmological production scenarios become possible, often with implications extending beyond DM to include dark radiation (DR), dark energy (DE), early dark energy, or even inflationary dynamics.

High-energy theories such as string theory naturally predict a plenitude of axion-like states, giving rise to the concept of the \emph{axiverse}~\cite{Arvanitaki:2009fg}. A caveat is in order when extrapolating the standard cosmological picture of axion production to string-theory realizations. In many extra-dimensional constructions, axions arise as zero modes of higher-dimensional gauge fields rather than as the pseudo–Nambu–Goldstone bosons of a spontaneously broken four-dimensional global $U(1)_{\rm PQ}$. In such cases, there is no conventional symmetry-breaking phase transition in the early Universe, and consequently the formation of a network of global cosmic strings is not generic. The standard post-inflationary scenario with axion strings and domain walls applies most directly to field-theoretic PQ axions. Cosmic strings in string-theory setups may still arise in more exotic cosmological frameworks, such as certain realizations of brane inflation, and can exhibit parametrically different tensions and dynamics when present~\cite{Benabou:2023npn}. In typical compactifications, the breaking of higher-dimensional gauge symmetries and the presence of antisymmetric tensor fields generate a large number of light pseudoscalars, each associated with distinct cycles of the compact manifold. Their masses and couplings are determined by non-perturbative effects such as instantons or gaugino condensation, leading to spectra that are exponentially hierarchical and broadly distributed across many decades~\cite{Witten:1984dg, Fox:2004kb, Conlon:2006tq, Svrcek:2006yi, Cicoli:2012sz, Demirtas:2018akl}.\footnote{See Ref.~\cite{Petrossian-Byrne:2025mto} for a recent realization of an axiverse populated in post-inflationary cosmologies.} As a result, the axiverse generically contains fields ranging from effectively massless ultralight ALPs, with Compton wavelengths comparable to galactic or even cosmological scales, up to heavier states that can behave as standard cold DM. This plenitude allows ALPs to participate in cosmology in diverse ways. For instance, ultralight ALPs with masses $m_a \sim 10^{-22}\,\mathrm{eV}$ act as DM, suppressing small-scale structure formation~\cite{Hu:2000ke, Hui:2016ltb, Marsh:2015xka}, while fields with $m_a \sim H_0$ evolve slowly on cosmological timescales and have been proposed as quintessence-like candidates for DE~\cite{Frieman:1995pm, Kaloper:2005aj,Nomura:2000yk, Visinelli:2018utg}.

Axions and ALPs can also play a role in the very early Universe, where axion-like fields have been proposed as inflaton candidates in models of natural inflation~\cite{Freese:1990rb, Adams:1992bn, Kim:2004rp, Dimopoulos:2005ac, Silverstein:2008sg, McAllister:2008hb, Visinelli:2011jy, Freese:2017ace}. Beyond their possible inflationary role, light ALPs with weak but non-negligible couplings to photons or fermions may remain in thermal contact with the plasma until MeV–GeV temperatures, subsequently contributing to DR and leaving imprints in the effective number of neutrinos $N_{\mathrm{eff}}$~\cite{Arias:2012az,Graf:2010tv,DEramo:2018vss,Caloni:2022uya,Salvio:2013iaa,Wantz:2009it,Archidiacono:2013cha, Barbieri:2026ewj}. In addition, heavier ALPs in the eV–keV range can decay into photons, producing diffuse backgrounds or line features testable by X-ray and $\gamma$-ray telescopes~\cite{Cadamuro:2011fd,Essig:2013goa,Conlon:2013txa,Grin:2006aw}. ALPs in the MeV range can inject energy into the primordial plasma and affect the reionization and thermal history of the Universe, with constraints arising from the cosmic microwave background (CMB)~\cite{Slatyer:2024nbm, Cheng:2025cmb, Yin:2025amn}, Big--Bang nucleosynthesis (BBN)~\cite{Millea:2015qra, Depta:2020wmr, Depta:2020zbh, Balazs:2022tjl}, and Lyman-$\alpha$ observations~\cite{Capozzi:2023xie}. Finally, axions and ALPs with masses in the GeV range are subject to direct searches at colliders and fixed-target experiments, where they may appear as missing energy signatures or displaced decays~\cite{Jaeckel:2015jla,Dolan:2017osp,Bauer:2017ris,Banerjee:2017hhz,Bernreuther:2019pfb}.

This versatility makes the axion framework a well-motivated portal connecting ultraviolet physics with infrared observables. The broad parameter space of the axiverse ties fundamental aspects of string compactifications to a wealth of phenomenology, ranging from black hole superradiance~\cite{Arvanitaki:2010sy,Baryakhtar:2017ngi,Blas:2020och} and inflationary reheating~\cite{Cicoli:2012cy,Acharya:2010zx}, to laboratory experiments probing axion–photon couplings~\cite{Sikivie:1983ip, Sikivie:1985yu, Raffelt:1987im, VanBibber:1987rq, Graham:2015ouw, Irastorza:2018dyq}. In this way, axions and ALPs serve as a rare bridge between the microphysics of quantum gravity and the macroscopic evolution of the Universe, yielding concrete, testable predictions across particle physics, astrophysics, and cosmology.

From a pedagogical perspective, axions also provide a unifying framework for introducing scalar field dynamics, quantum field theory in the early Universe, and the interplay between astrophysics and particle physics. They illustrate how a single class of particles can address seemingly disparate problems:
\begin{itemize}
\item As \textbf{dark matter}~\cite{Sikivie:2006ni, DiLuzio:2020wdo, Arza:2026rsl}, axions are produced non-thermally via the misalignment mechanism~\cite{Preskill:1982cy, Abbott:1982af, Dine:1982ah} and the decay of topological defects~\cite{Davis:1986xc, Harari:1987ht}. Their wave-like properties on small scales may also alleviate tensions with the standard $\Lambda$CDM paradigm~\cite{Kim:2008hd,Hu:2000ke,Hui:2016ltb}.
\item As \textbf{dark energy}, ultralight ALPs with masses $m_a \sim H_0$ can act as dynamical quintessence fields, providing an alternative to a cosmological constant~\cite{Frieman:1995pm, Kaloper:2005aj,Nomura:2000yk,Visinelli:2018utg}.
\item As \textbf{dark radiation}, axions produced thermally or through the decays of heavy fields contribute to the relativistic energy density, modifying the effective number of neutrino species $N_{\mathrm{eff}}$~\cite{Mangano:2005cc,Fields:2019pfx,Arias:2012az,DEramo:2018vss}.
\end{itemize}
Axions unite high-energy theory, cosmology, astrophysics, and observations, making them a well-motivated Standard Model extension and a case study of physics across scales.

In the following sections, we review selected aspects of these cosmological roles. Section~\ref{sec:DM} discusses axions as DM, including production mechanisms and constraints. Section~\ref{sec:DE} explores axions as DE, focusing on quintessence. Section~\ref{sec:DR} presents axions as DR and implications for $N_{\mathrm{eff}}$. We conclude in Section~\ref{sec:conclusions} with perspectives on the role of axions in upcoming theoretical and observational programs.

\section{Axions as Dark Matter}
\label{sec:DM}

\subsection{Motivations}

The most extensively studied role of the axion is as a DM candidate. A defining feature is that axions are produced non-thermally in the early Universe, leading to a phase-space distribution characteristic of cold, pressureless matter. The canonical production channel is the \emph{misalignment mechanism}. At early times, when the Hubble expansion rate $H$ exceeds the temperature-dependent axion mass $m_a(T)$, the axion field $a$ remains frozen at some initial misalignment angle $\theta_i$ set by Peccei--Quinn symmetry breaking. Once the expansion rate drops below the axion mass, the field begins coherent oscillations about the minimum of its potential. These oscillations redshift as nonrelativistic matter, naturally giving rise to a relic abundance. A simple estimate yields
\begin{equation}
    \Omega_a h^2 \simeq 0.12 \left( \frac{f_a}{10^{12}\,\text{GeV}} \right)^{7/6}\theta_i^2\,,
\end{equation}
where $f_a$ is the Peccei–Quinn scale, and the exponent $7/6$ arises from the temperature dependence of the axion mass around the QCD epoch~\cite{Preskill:1982cy,Abbott:1982af,Dine:1982ah,Turner:1985si}.

This standard picture has been refined in several important directions. First, anharmonic corrections to the axion potential, relevant when the initial misalignment angle is large, delay the onset of oscillations. This leads to an enhancement of the relic abundance compared to the naive quadratic approximation~\cite{Visinelli:2009zm}. Second, modifications to the cosmological history can leave a significant imprint on the relic density. If the Universe underwent periods of early matter domination, kination, or late entropy release, the Hubble rate at the time of axion oscillations would differ from the standard radiation-dominated case. This shifts the onset of oscillations and hence alters the final abundance~\cite{Grin:2007yg,Hertzberg:2008wr,Visinelli:2009kt,Wantz:2009it,Visinelli:2014twa,Arias:2012az,Allahverdi:2020bys}. A more recent development is the \emph{kinetic misalignment mechanism}, in which the axion field begins with a nonzero initial velocity. In this case, the relic abundance can be dramatically modified, with a much weaker dependence on the initial misalignment angle $\theta_i$~\cite{Co:2019jts,Co:2019wyp,Barman:2021rdr}. Additional modifications occur when axions couple to dark sectors or experience frictional effects in the plasma. Such interactions can delay the onset of oscillations, once again modifying the relic abundance~\cite{Papageorgiou:2022prc,Agrawal:2022lsp}. Similarly, if the axion field starts close to the maximum of its potential, a situation known as \emph{hilltop initial conditions}, the delayed onset of oscillations strongly enhances the relic density~\cite{Lyth:1991bb,Visinelli:2017imh}. The role of reheating has also been scrutinized in detail. If reheating is not instantaneous, or if late-decaying fields dominate the energy budget, the effective temperature at which oscillations begin is shifted. This can substantially modify the viable axion mass window~\cite{Chang:1998ys, Bae:2008ue, Bae:2017hlp, Blum:2014vsa, Harigaya:2019tzu, DiLuzio:2021gos, Co:2019wyp, Visinelli:2018wza, Mazde:2022sdx}. Finally, a variety of models have explored scenarios with non-standard axion mass generation or additional PQ breaking effects. In these cases, the temperature dependence of the mass or the structure of the axion potential can deviate significantly from the standard QCD prediction, opening qualitatively new possibilities for axion DM and affecting the predictivity of the relic abundance~\cite{Nakayama:2021avl, Zantedeschi:2026iql}.

\subsection{Numerical studies}

A crucial ingredient in axion cosmology is the temperature dependence of the axion mass, $m_a(T)$, which is directly related to the topological susceptibility of QCD. At low temperatures, chiral perturbation theory provides a reliable description, but around and above the QCD crossover, nonperturbative input is essential. For many years, phenomenological models such as the dilute instanton gas approximation (DIGA) were used to estimate the high-temperature behavior, but these carry large uncertainties. This motivated high-precision lattice QCD computations of the topological susceptibility, which now provide the state-of-the-art determination of the axion mass across the relevant thermal range~\cite{Borsanyi:2016ksw, Petreczky:2016vrs, Bonati:2016tvi, Burger:2018fvb, Lombardo:2020bvn, Kotov:2021rah, Borsanyi:2022xml}. These results sharpen relic density predictions from the misalignment mechanism and enable global fits that refine the preferred axion mass range~\cite{Buschmann:2021sdq, Gorghetto:2020qws, Arias:2012az, DiLuzio:2020wdo}. While misalignment remains the most robust production channel, the quantitative relic abundance is highly sensitive to both microphysics and cosmological history.

Despite this robust qualitative picture, quantitatively predicting the axion relic abundance from defect decay remains one of the central open problems in numerical cosmology. The challenge arises because the relevant dynamics couple microscopic and cosmological scales: the string core width is many orders of magnitude smaller than the Hubble radius, making it impossible to resolve both simultaneously. Consequently, numerical studies must adopt approximations for the unresolved core structure and extrapolate across many decades of scale separation. Early works~\cite{Battye:1994au, Chang:1998tb, Hiramatsu:2010yn} established the basic phenomenology, but more recent large-scale simulations have reached divergent conclusions. Some groups find that axion emission is dominated by soft, low-momentum modes, implying a large enhancement of the relic density and favoring relatively heavy axions in the $\mathcal{O}(100)\,\mu{\rm eV}$ range~\cite{Hiramatsu:2012gg, Gorghetto:2018myk, Gorghetto:2020qws, Gorghetto:2024vnp, Kaltschmidt:2024uxt}. In contrast, other sophisticated approaches, which employ improved treatments of the string core or adaptive mesh refinement, argue that the radiation spectrum is significantly harder, reducing the infrared enhancement of the axion number density and leading to a smaller total relic abundance than in simulations that find predominantly soft emission~\cite{Klaer:2017ond, Buschmann:2021sdq, Buschmann:2024bfj}. A larger contribution to the axion density may arise when the collapse of the string domain wall network is taken into account~\cite{Benabou:2024msj}. In those string-theory realizations in which axion strings do form, their tensions can be parametrically larger than those of field-theory strings, potentially altering their cosmological impact~\cite{Benabou:2023npn}. In string-theory UV completions that incorporate a GUT with gauge coupling unification at the supersymmetric GUT scale, the mass of the QCD axion can be bounded from above by $10^{-8}$\,eV, independently of the DM fraction it constitutes~\cite{Benabou:2025kgx}. The discrepancies in DM axion mass predictions reflect sensitivities to technical issues such as the long-term scaling behavior of the string network~\cite{Kawasaki:2018bzv}, the inclusion of loop versus long-string populations~\cite{Davis:1989nj, Saikawa:2012uk}, and the treatment of possible logarithmic corrections to the string tension~\cite{Gorghetto:2018myk}. More exotic dynamics, such as superconducting strings~\cite{Witten:1984eb}, may also influence energy-loss channels. Resolving this tension is essential: the axion relic density from defects directly fixes the preferred DM mass range and guides the targets of ongoing and future experimental searches.

\subsection{Topological defect decay}

If the Peccei–Quinn (PQ) symmetry is broken after inflation, the axion field is not homogenized across the observable Universe and topological defects inevitably form in field-theoretic PQ models. In particular, global cosmic strings emerge at the PQ-breaking scale and persist until the QCD epoch, when the axion potential turns on and domain walls appear. The subsequent evolution of this string--wall network leads to copious axion emission, providing an additional, often dominant, contribution to the relic abundance beyond the misalignment mechanism. The quantitative outcome depends sensitively on the efficiency of axion radiation from the string network, an issue under active investigation through numerical simulations and analytic modeling~\cite{Kawasaki:2014sqa, Hiramatsu:2012gg, Gorghetto:2018myk, Gorghetto:2020qws, Klaer:2017ond, Buschmann:2019icd, Vaquero:2018tib, Buschmann:2024bfj, Buschmann:2021sdq, Witten:1984eb, Benabou:2024msj}. Despite significant progress, the relative weights of misalignment and defect decay in setting the relic density remain under debate.

An additional layer of complexity arises from the domain wall number $N_{\rm DW}$. For $N_{\rm DW}=1$, the string–wall network is unstable and decays efficiently, releasing a burst of axions. For $N_{\rm DW}>1$, however, the network is stable and would eventually dominate the energy density unless explicitly destabilized, typically by a small PQ--breaking operator. The details of this decay crucially affect the final relic abundance and may shift the preferred axion mass window by more than an order of magnitude. Besides these non-thermal channels, axions and ALPs can also be produced thermally through interactions with gluons, photons, and fermions. While typically subdominant for the QCD axion, thermal production can dominate for heavier or more strongly coupled ALPs, yielding distinctive cosmological and astrophysical signatures~\cite{Masso:2002np, Salvio:2013iaa, Bolliet:2020ofj}.

The phenomenology resulting from these mechanisms is remarkably rich. In the canonical QCD axion window, $m_a \sim (10^{-5}$--$10^{-3})$\,eV, axions behave as cold DM consistent with large-scale structure. At much lower masses, $m_a \lesssim 10^{-21}$\,eV, they act as ultralight DM with kiloparsec-scale de Broglie wavelengths that suppress small-scale structure. This has been proposed as a remedy to the small-scale challenges of $\Lambda$CDM~\cite{Hu:2000ke, Arvanitaki:2009fg, Schive:2014dra, Schive:2014hza, Hui:2016ltb, Irsic:2017yje, Veltmaat:2018dfz, Rogers:2020ltq, Nori:2018pka}, though Lyman-$\alpha$ data place strong lower bounds on the mass. Furthermore, the inhomogeneous initial conditions and non-linear dynamics of axion overdensities can give rise to compact clumps, or miniclusters, with possible substructures known as axion stars. These objects form naturally from string-induced fluctuations or dynamics at the QCD phase transition, and may be probed via microlensing, pulsar timing, or transient signatures in direct detection experiments~\cite{Hogan:1988mp, Kolb:1993hw, Kolb:1993zz, Duffy:2008dk, Fairbairn:2017dmf, Visinelli:2017ooc, Schiappacasse:2017ham, Visinelli:2018wza, Eggemeier:2019khm, Eggemeier:2020zeg, Ellis:2020gtq, Yin:2024xov}.

\subsection{Bounds from astrophysics and cosmology}

Astrophysical and cosmological probes provide some of the most stringent constraints on axions and ALPs. Stellar evolution is highly sensitive to additional cooling channels, such as axion emission via the Primakoff process or axion–nucleon bremsstrahlung. Observations of horizontal branch stars, red giants, and white dwarfs have been used to bound the axion–photon and axion–electron couplings~\cite{Raffelt:2006cw, Viaux:2013lha, Capozzi:2020cbu, Carenza:2020cis, DiLuzio:2020jjp}. Core-collapse supernovae provide complementary constraints, most notably from SN1987A, which restricts excessive energy loss through axion emission~\cite{Raffelt:1987yt, Raffelt:2006cw, Carenza:2019pxu, Caputo:2024oqc}. On cosmological scales, axions with masses below around the eV remain stable on cosmic timescales and can act as DM or DR~\cite{Arias:2012az, Marsh:2015xka}. For ultralight axions, the suppression of structure formation leaves an imprint on the matter power spectrum and on the CMB lensing signal~\cite{Hu:2000ke, Hlozek:2016lzm, Hlozek:2017zzf, Poulin:2018dzj, Rogers:2020ltq, Hui:2021tkt, Ferreira:2020fam}. Recent high-redshift observations from the James Webb Space Telescope further probe this regime by constraining early galaxy formation and the ultraviolet luminosity function, placing additional bounds on ultralight axion DM scenarios~\cite{Bird:2023pkr, Lazare:2024uvj, Winch:2024mrt, Sipple:2024svt}. Current large-scale structure and CMB data place strong upper limits on the abundance of axions with $m_a \lesssim 10^{-24}$\,eV. Furthermore, if the PQ symmetry is broken before inflation, isocurvature fluctuations in the axion field are generated, whose amplitude is constrained by {\it Planck} data~\cite{Linde:1991km, Seckel:1985tj, Kawasaki:2013ae}. These isocurvature bounds translate into strong limits on the inflationary Hubble scale and on the allowed axion parameter space.

A further powerful probe arises from black hole superradiance. If the axion Compton wavelength matches the gravitational radius of astrophysical black holes, superradiant instabilities lead to the extraction of rotational energy, forming large bosonic clouds. The absence of highly spinning black holes in certain mass ranges translates into constraints on axions with masses $m_a \sim (10^{-20}$--$10^{-10})$\,eV~\cite{Arvanitaki:2010sy, Yoshino:2013ofa, Brito:2015oca, Berti:2015itd}. Future gravitational wave observatories may directly detect the continuous radiation emitted by these clouds, providing a novel observational window~\cite{Arvanitaki:2016qwi, Sprague:2024lgq}. Additional signatures may arise if axions couple directly to gravity through parity-violating operators, leading to monochromatic gravitational wave signals from axion DM decay~\cite{Figliolia:2025dtw}.

Complementary to these indirect bounds, radio telescopes offer the possibility of directly searching for axion DM in the Galactic halo. Axion–photon conversion in astrophysical magnetic fields, or axion decay into two photons, can give rise to narrow spectral features in the radio band, with a linewidth set by the Galactic velocity dispersion. This makes radio facilities particularly well-suited to ultra-narrow line searches~\cite{Lai:2006af, Pshirkov:2007st}. Observations with the Green Bank Telescope (GBT), the Effelsberg 100m telescope, and Low Frequency Array (LOFAR) have been employed to place limits on the axion–photon coupling across a wide mass range~\cite{Safdi:2018oeu, Foster:2020pgt, Darling:2020uyo, Nurmi:2021xds, Foster:2022fxn, Noordhuis:2022ljw}. In parallel, axion substructures can boost the search in radio transients~\cite{Edwards:2020afl, Walters:2024vaw, Bhura:2026bpy}. These searches have boosted the theoretical modeling of the axion--photon conversion in astrophysical magnetic fields~\cite{Witte:2021arp, Witte:2022cjj, Gines:2024ekm} and the distribution of galactic DM structures~\cite{Kavanagh:2020gcy, Xiao:2021nkb, DSouza:2024flu, DSouza:2024uud}. Looking ahead, next-generation observatories such as the Square Kilometre Array (SKA) are expected to extend the sensitivity to axion masses in the $\mu$eV–neV range.

\subsection{Laboratory searches}

The modern program of laboratory searches for cosmic axions originates from the seminal work of Pierre Sikivie, which established the conversion of axions into photons in strong magnetic fields as a viable detection strategy, laying the foundation for resonant cavity haloscope experiments~\cite{Sikivie:1983ip, Sikivie:1985yu, Raffelt:1987im, VanBibber:1987rq, Raffelt:1996wa}. This proposal led to the first haloscope experiments at Brookhaven National Laboratory and the University of Florida, where a resonant microwave cavity was placed inside a superconducting magnet to search for the narrowband photon signal expected from galactic axion DM~\cite{Hagmann:1990tj}. Although these pioneering searches were limited in sensitivity, they established the foundation for a worldwide experimental effort. The most advanced realization of this concept is the Axion Dark Matter eXperiment (ADMX), which has achieved sensitivity to realistic QCD axion models in the $\mu$eV mass range~\cite{ADMX:2018gho, Bartram:2021ysp}. ADMX employs a high-$Q$ tunable microwave cavity immersed in an 8~T magnetic field, combined with ultra-low-noise quantum amplifiers that approach the standard quantum limit. By scanning over cavity resonances, ADMX is systematically probing the classic KSVZ and DFSZ benchmark models, and its successive upgrades are steadily extending the explored parameter space. The QUest for AXion (QUAX) program within the Istituto Nazionale di Fisica Nucleare (INFN) pursues two Sikivie-type haloscopes, with experiments operating at Legnaro~\cite{QUAX:2025wtb} and at LNF~\cite{QUAX:2024fut}. The Legnaro search, centered near $42\,\mu$eV, excludes axion–photon couplings above $10^{-14}\,\mathrm{GeV}^{-1}$~\cite{QUAX:2025wtb}, while the LNF search in the $(36.52413$--$36.5511)\,\mu$eV window sets a limit of $g_{a\gamma} < 0.882\times 10^{-13}\,\mathrm{GeV}^{-1}$ at 90\% confidence level~\cite{QUAX:2024fut}.

Other cavity-based searches, targeting diverse frequency windows, are coming online in the near future. FLASH (the FINUDA magnet for Light Axion SearcH) is conceived as a large resonant-cavity haloscope to be operated in the cryogenic FINUDA magnet at the Laboratori Nazionali di Frascati (LNF) near Rome, and is designed to probe the previously underexplored low-frequency window around $\sim(0.5$–$1.5)\,\mu$eV using resonant enhancement and a dedicated quantum-limited readout chain (microstrip SQUID amplifiers)~\cite{Alesini:2019nzq, Alesini:2023qed}. Similar schemes are being developed by the Taiwan Axion Search Experiment with Haloscope (TASEH), targeting the (4.7–4.8)\,GHz band~\cite{TASEH:2022noe}; by the Axion Longitudinal Plasma HAloscope (ALPHA), which is pursuing a novel tunable cryogenic plasma haloscope sensitive to axion masses matching the plasma frequency~\cite{ALPHA:2022rxj}; and by RADES (CERN Axion Search with High-Temperature Superconducting Cavity), operating at 8.84\,GHz~\cite{Ahyoune:2024klt}. Complementary strategies have been devised to reach higher mass ranges, where conventional cylindrical cavities become prohibitively small. Dielectric haloscopes, exemplified by the MADMAX concept, employ large-area stacks of dielectric disks to coherently enhance axion–photon conversion and are aimed at the tens-to-hundreds of $\mu$eV window~\cite{Caldwell:2016dcw, Millar:2016cjp}. At the opposite end, the DMRadio-m3 experiment is designed to probe the QCD axion mass range below $1\,\mu$eV using a static magnetic field $\gtrsim 4\,$T, a coaxial inductive pickup, a tunable LC resonator, and a DC-SQUID readout~\cite{DMRadio:2022pkf}.

Other experimental efforts target couplings beyond the axion–photon portal. Within the QUAX collaboration, the QUAX$_{\rm ae}$ program searches for the axion–electron interaction by exploiting electron-spin resonances in magnetized materials~\cite{DellaValle:2014xoa, Barbieri:2016vwg}. In this scenario, the nonrelativistic axion field acts as an effective oscillating radio-frequency magnetic field that can excite magnon modes in a polarized sample. Practically, QUAX embeds a magnetized sample inside a microwave cavity or readout circuit, tunes the static magnetic field so that the electron-spin Larmor frequency coincides with the axion frequency, and detects the resulting magnetization oscillations with ultra-low-noise amplifiers or SQUIDs. Hybridization between cavity and magnon modes is further employed to enhance signal coupling. Prototype runs have already set limits on the axion–electron coupling strength at mass $\simeq 58\,\mu$eV, and the tunable QUAX program has since reported extended results~\cite{Crescini:2018qrz}. This detection channel is particularly well matched to DFSZ-like models with unsuppressed axion–electron couplings and benefits from continuing advances in material engineering and quantum-limited readout.

By contrast, NMR-based concepts such as CASPEr target axion–nucleon couplings (and axion-induced oscillating nuclear electric dipole moments). CASPEr implements precision magnetometry / comagnetometry and resonant NMR detection schemes (CASPEr–Wind and CASPEr–Electric variants) to search for tiny, coherent oscillatory signals from the ambient axion field over very low masses; these techniques are complementary in coupling and in target mass range to QUAX and photon-based haloscopes~\cite{Budker:2013hfa}.

Additional complementary approaches include light-shining-through-a-wall (LSW) experiments, which provide purely laboratory-based probes of axion–photon mixing without relying on astrophysical or cosmological assumptions. In these setups, a laser beam traverses a strong magnetic field where photons may convert into axions, which can then pass through an opaque barrier and reconvert into photons in a second magnetic field. While early pioneering efforts established the first laboratory bounds on the axion–photon coupling~\cite{Ruoso:1992nx}, more recent LSW initiatives continue to expand the search. The Any Light Particle Search (ALPS) experiment at DESY has pushed sensitivities into previously unexplored regions of parameter space~\cite{Ehret:2010mh}, and ALPS-II, currently under construction, will employ long-baseline optical cavities to enhance sensitivity by several orders of magnitude relative to its predecessors~\cite{Bahre:2013ywa}. Bounds on heavy axions with mass within $[10^{-3}, 10^{4}]$\,eV have been placed by the European X-ray Free Electron Laser facility
(EuXFEL)~\cite{Halliday:2024lca}.

Another powerful laboratory technique is the use of helioscopes, which search for axions produced in the core of the Sun and subsequently reconverted into X-rays in a strong laboratory magnetic field. The CAST experiment at CERN has set world-leading bounds on the axion–photon coupling over a wide mass range~\cite{CAST:2007jps, CAST:2017uph}, and the proposed next-generation International Axion Observatory (IAXO) is expected to improve sensitivity by more than an order of magnitude~\cite{Irastorza:2011gs, IAXO:2019mpb}. The solar axion flux is dominated by production through the Primakoff process, $\gamma + N \to N + a$, with additional contributions from Compton-like scattering, $\gamma + e \to a + e$, and electron–nucleus bremsstrahlung, $e + N \to e + N + a$. A key astrophysical point is that stellar energy-loss rates are limited by the rate at which stars can radiate energy away. Weakly coupled light bosons, such as axions, provide an efficient escape channel since they leave the stellar plasma unimpeded, in contrast to photons, which undergo repeated scattering and are emitted only from the stellar surface. These arguments underlie classical stellar cooling bounds on axion couplings~\cite{Raffelt:2006cw, Viaux:2013lha, Capozzi:2020cbu, Caputo:2024oqc}, and helioscopes exploit the same physical processes in a controlled laboratory setting. A dedicated next step in helioscope searches is the babyIAXO prototype, designed as a pathfinder for the full IAXO. BabyIAXO will employ a smaller-scale magnet and detector system to validate the technological concepts and detection strategies, while already probing new parameter space for solar axions~\cite{IAXO:2019mpb}. Future concepts further exploit high-intensity photon sources from free-electron lasers, targeting ALPs in the sub–THz frequency range with laboratory-based setups such as the Sub-THz Axion eXperiment (STAX)~\cite{Graham:2015ouw}.

At even higher frequencies, dish antenna concepts provide a broadband method to search for axions and axion-like particles. In this approach, galactic DM axions convert into photons at the surface of a conducting dish placed in a strong magnetic field, producing a faint but coherent electromagnetic wave that can be detected by sensitive receivers~\cite{Horns:2012jf}. Because no resonant structure is required, dish antennas naturally cover wide mass ranges in a single setup, at the cost of reduced sensitivity compared to resonant cavities. Complementing these laboratory-based approaches, broadband radio observatories such as the Five-hundred-meter Aperture Spherical Telescope (FAST)~\cite{2011IJMPD..20..989N}, LOFAR~\cite{2013A&A...556A...2V}, and SKA~\cite{1999sska.conf.....T} are being explored as probes of axion-induced radio signals. These could arise either from DM conversion in astrophysical magnetic fields or from transient phenomena such as axion minicluster encounters~\cite{Sigl:2017sew, CAST:2017uph}. Together, these approaches extend the experimental program into frequency ranges that are difficult to reach with conventional resonant techniques, highlighting the diversity of strategies being pursued to explore the full axion parameter space. Collectively, these experimental strategies now cover many decades in axion mass and coupling. The field has evolved from a single theoretical proposal into a broad program that probes well-motivated regions of the QCD axion and ALP parameter space.

\section{Axions as Dark Energy}
\label{sec:DE}

We restrict the discussion to canonical axion quintessence and do not attempt to review the broader landscape of dynamical DE models. A qualitatively distinct role for axions arises if their mass is comparable to the present-day Hubble scale, $m_a \sim H_0 \sim 10^{-33}\,\text{eV}$. In this regime, the axion field evolves so slowly that it can effectively behave as a form of dynamical dark energy, or quintessence~\cite{Ratra:1987rm,Wetterich:1987fm,Caldwell:1997ii}. The standard potential takes the form
\begin{equation}
    V(a) = \Lambda^4 \left[1 - \cos\left(\frac{a}{f_a}\right)\right]\,,
\end{equation}
with an effective mass given by $m_a \sim \Lambda^2/f_a$. When displaced from the minimum, the field undergoes slow-roll evolution governed by
\begin{equation}
    3H\dot{a} + V'(a) \simeq 0\,,
\end{equation}
leading to an effective equation-of-state parameter
\begin{equation}
    w_{\rm DE}(z) \simeq -1 + \frac{\dot{a}^2}{V(a)}\,,
\end{equation}
which remains close to $-1$, but can deviate in a time-dependent manner depending on the shape of the potential and the initial misalignment~\cite{Frieman:1995pm, Caldwell:1997ii, Peebles:1998qn, Steinhardt:1999nw}.

Axion quintessence has also been discussed in connection with the ``swampland conjectures,'' which argue that not all effective field theories admit consistent embeddings in quantum gravity~\cite{Obied:2018sgi, Agrawal:2018own}. The de~Sitter swampland conjecture, in particular, disfavors scalar potentials that support stable or metastable de~Sitter vacua. Axion models, by contrast, feature periodic or runaway potentials protected by a shift symmetry, allowing them, in suitable constructions, to be compatible with these constraints. This makes axion quintessence a particularly compelling DE candidate in string-inspired frameworks~\cite{Cicoli:2018kdo, Achucarro:2018vey, Heisenberg:2018yae}. A further advantage is that the approximate shift symmetry protects the flatness of the potential against radiative corrections, unlike in generic scalar-field models. This property helps resolve the naturalness problem of sustaining an ultralight scalar degree of freedom stable over cosmological timescales~\cite{Frieman:1995pm, Choi:1999xn, Panda:2010uq, Gupta:2015uea}.

This symmetry protection is central to the appeal of axion quintessence. Generic quintessence fields require an ultralight mass, $m\sim H_0$, which is typically unstable against quantum corrections. Axions circumvent this issue because the approximate shift symmetry suppresses dangerous contributions to the potential. In ultraviolet completions such as the string ``axiverse,'' an exponentially large spectrum of axion masses can populate the ultralight window around the Hubble scale, making quintessence a natural outcome of high-energy model building~\cite{Arvanitaki:2009fg, Choi:1999xn, Kim:2002tq, Kim:2003qt, Svrcek:2006yi, Acharya:2010zx}. This protection, together with the possibility of consistent embedding in string compactifications, provides strong motivation for considering axions as DE candidates.

At the level of background cosmology, a slowly rolling axion modifies the Hubble expansion rate $H(z)$, leading to departures from $\Lambda$CDM in distance-based probes such as Type Ia supernovae and baryon acoustic oscillations (BAO). Dynamical DE also affects the growth of large-scale structure, leaving imprints in redshift-space distortions and weak gravitational lensing. These signatures are most pronounced at late times ($z\lesssim 2$), when DE becomes dynamically relevant, making low-redshift surveys especially sensitive to axion quintessence. On perturbative scales, canonical quintessence fields, including axions with standard kinetic terms, exhibit a relativistic sound speed, $c_s^2\simeq 1$, which suppresses clustering below the horizon and renders the DE effectively smooth on galactic scales. This behavior sets axion quintessence apart from other dark sector models (e.g., coupled DE, clustering DE, or certain modified-gravity scenarios) in which spatial inhomogeneities can be significant. Even so, a time-evolving component alters the gravitational potential on large scales, producing an integrated Sachs–Wolfe (ISW) contribution to the cosmic microwave background. Since the ISW effect is sensitive to the time variation of gravitational potentials, it provides an independent probe of dynamical DE complementary to geometric and growth-based observables~\cite{Calabrese:2010uf,Ade:2015rim}.

Current cosmological datasets constrain the DE equation of state to lie within a few percent of $w=-1$, allowing only mild time dependence compatible with axion potentials~\cite{Dutta:2009dr,Hlozek:2014lca,Marsh:2015xka,Choi:2016kke}. The next generation of surveys will sharpen these bounds significantly. Euclid and the Vera Rubin Observatory (LSST) will deliver precise measurements of galaxy clustering, weak lensing, and supernova distances across wide redshift ranges~\cite{Amendola:2012ys,Ivezic:2008fe}. The Nancy Grace Roman Space Telescope will extend and calibrate supernova Hubble diagrams and BAO measurements~\cite{Spergel:2015sza}, while upcoming CMB experiments such as Simons Observatory and LiteBIRD will tighten constraints on the ISW effect and CMB lensing~\cite{Abazajian:2019eic, LiteBIRD:2022cnt, SimonsObservatory:2018koc}. Collectively, these observations are expected to probe percent-level deviations from a pure cosmological constant in the equation of state $w_{\rm DE}(z)$, with target sensitivities of order $\Delta w\sim 0.01$. They will also test characteristic correlations between geometry and growth that are distinctive to dynamical DE.

Beyond single-field setups, interactions among multiple ultralight axions give rise to richer phenomenology. Multi-field dynamics can produce thawing or oscillatory equations of state, tracking behavior, transient features in the expansion history, and nontrivial energy exchange between sectors. In axiverse scenarios, the same high-energy completion that yields an ultralight quintessence axion often also provides heavier axions acting as DM. This makes it essential to search for consistent signatures across cosmology and astrophysics~\cite{Das:2005yj, Cicoli:2012sz, Gupta:2015uea, Agrawal:2018own, Kaloper:2019xfj, Brandenberger:2020gaz,Foster:2022ajl, Petrossian-Byrne:2025mto}. Linking expansion history, structure growth, DM searches, and small-scale astrophysical probes shows the importance of coordinated observational strategies to test axion-based explanations of cosmic acceleration.

On the theory side, axion quintessence raises important questions about quantum gravity constraints. Arguments based on the weak gravity conjecture and various swampland criteria may place limits on field excursions, decay constants, or the structure of the potential, with direct consequences for the viability of large-field quintessence constructions~\cite{Rudelius:2015xta, Brown:2015iha, Ooguri:2018wrx, Agrawal:2018own}. However, targeted model-building and explicit constructions in string compactifications demonstrate that axion quintessence can be realized in controlled setups consistent with UV expectations; radiative stability, shift-symmetry protection, and the availability of multiple axions are key ingredients to these constructions~\cite{Heisenberg:2018rdu, Akrami:2020zxw, Gonzalo:2021zsp, Cicoli:2021gss, Mehta:2021pwf}. Recent work suggests that realistic axion potentials can produce observationally acceptable quintessence dynamics without extreme fine-tuning~\cite{Lin:2025gne}.

Finally, an active line of research investigates if axion quintessence can alleviate cosmological tensions. Slowly rolling or transiently evolving axion fields have been proposed as possible mechanisms to address the $H_0$ tension and certain late-time anomalies in large-scale structure, with a number of phenomenological analyses and explicit model constructions appearing in the literature~\cite{Poulin:2018dzj, Agrawal:2019lmo, DiValentino:2019exe, Ye:2020btb, Choi:2021nql, DiValentino:2021izs, Urena-Lopez:2025rad, Luu:2025fgw}. However, this possibility is subject to important limitations. In particular, Ref.~\cite{Colgain:2025nzf} shows that canonical quintessence models with $w(z)>-1$ generically predict a reduction of the inferred $H_0$ relative to $\Lambda$CDM, worsening the Hubble tension. A complementary analysis based on low-redshift reconstructions of the expansion history also indicates that quintessence dynamics are typically disfavored as a solution to this tension~\cite{Banerjee:2020xcn}. While axion quintessence may still play a role in addressing other late-time anomalies, this possibility is highly model-dependent; moreover, current evidence disfavors it as an explanation of the $H_0$ discrepancy. The scenario nevertheless remains an attractive and testable alternative to a cosmological constant, owing to its theoretical motivation and the observational reach of future surveys.

\section{Axions as Dark Radiation}
\label{sec:DR}

If axions couple to SM fields, they can be produced abundantly in the early Universe and contribute to the radiation density as extra relativistic degrees of freedom~\cite{ParticleDataGroup:2024cfk}. This effect is conventionally parametrized by the effective number of neutrino species, $N_{\mathrm{eff}}$, whose departure from the SM prediction signals the presence of DR.

Axions may be produced thermally via interactions with quarks, gluons, charged leptons and photons. For the QCD axion, the axion–gluon coupling follows from the Peccei–Quinn anomaly, while its low-energy effective interactions, including derivative couplings to fermions and couplings to photons, are described by the general axion effective Lagrangian~\cite{Georgi:1986df, GrillidiCortona:2015jxo} and arise in ultraviolet completions such as the KSVZ and DFSZ frameworks~\cite{Kim:1979if, Shifman:1979if, Zhitnitsky:1980tq,Dine:1981rt}. Thermalization in the early Universe proceeds through processes such as $q g \leftrightarrow q a$, $g g \leftrightarrow g a$ and, at lower temperatures, electron/photonic channels like $e^\pm \gamma \leftrightarrow e^\pm a$ and $\gamma\gamma\leftrightarrow a a$~\cite{Turner:1986tb, Srednicki:1985xd, Masso:2002np, Chang:1993gm, Brandenburg:2004du, Graf:2010tv, Millea:2015qra, Bouzoud:2024bom}. The Primakoff process ($\gamma q \to q a$ or $\gamma e\to e a$) in a charged plasma is often an efficient thermal production channel and is controlled by the axion–photon coupling $g_{a\gamma}$. At still lower temperatures axions are produced in atomic, molecular, and nuclear transitions; these late-time processes matter for astrophysical searches but are negligible for the cosmological DR budget compared to early thermal production.

If an axion species was once in full thermal equilibrium and decoupled at temperature $T_{\mathrm{dec}}$, entropy conservation fixes the present axion-to-photon temperature ratio,
\begin{equation}
    \frac{T_a}{T_\gamma} \;=\; \left(\frac{g_{*s}(T_\gamma)}{g_{*s}(T_{\mathrm{dec}})}\right)^{1/3},
\end{equation}
where $g_{*s}$ is the effective number of entropy degrees of freedom. The contribution to the relativistic energy density is then
\begin{equation}
    \Delta N_{\mathrm{eff}} \;=\; \frac{4}{7}\left(\frac{T_a}{T_\nu}\right)^4 ,
\end{equation}
with $T_\nu=(4/11)^{1/3}T_\gamma$. For example, decoupling before the QCD crossover, corresponding to $g_{*s}(T_{\mathrm{dec}})\sim \mathcal{O}(60)$, yields $\Delta N_{\mathrm{eff}}\simeq 0.03$~\cite{Cadamuro:2010cz, Wantz:2009it, Graf:2010tv, Melchiorri:2007cd}. This represents the minimal thermal relic contribution for axions decoupling above the QCD scale and constitutes a well-motivated target for next-generation CMB experiments. Conversely, later decoupling, e.g.\ after QCD confinement or after $e^\pm$ annihilation, leads to larger temperature ratios and can increase $\Delta N_{\mathrm{eff}}$ to $\gtrsim 0.2$.

A useful distinction is between the QCD axion and generic ALPs. For the QCD axion the mass and decay constant are related approximately by $m_a f_a\sim m_\pi f_\pi$, which ties couplings to the axion mass and makes thermalization efficient for sufficiently low $f_a$, e.g.\ $f_a\lesssim 10^{8}\text{--}10^{9}\,$GeV, depending on the dominant coupling~\cite{Turner:1986tb, Salvio:2013iaa}. ALPs instead allow decoupled mass and coupling scales, opening parameter space for very light, thermally-produced DR with sizable $g_{a\gamma}$ but sub-eV masses~\cite{Arias:2012az}.

Axions/ALPs may also be created non-thermally. A common realization in string and moduli cosmology is the decay of heavy moduli, saxions or inflatons into relativistic axions, which produces a DR component quantified by
\begin{equation}
    \Delta N_{\mathrm{eff}} \;=\; \frac{8}{7}\left(\frac{11}{4}\right)^{4/3}\frac{\rho_a}{\rho_\gamma},
\end{equation}
where $\rho_a$ is the axion energy density injected by decays. Depending on reheat temperature and branching ratios, $\Delta N_{\mathrm{eff}}$ from such decays can approach ${\cal O}(1)$ and is often constrained by CMB and BBN observations~\cite{Planck:2018vyg, Pandey:2018wvh, DEramo:2018vss, Caloni:2022uya, Badziak:2024szg, Barbieri:2026ewj}. In addition to the decay-driven DR studied in the moduli literature, recent work emphasises an \emph{irreducible} freeze-in population of axions arising from minimal couplings to SM fields; this population can produce cosmologically relevant axion abundances and leads to novel constraints in portions of parameter space~\cite{Langhoff:2022bij}.

Accounting for SM corrections, such as QED effects and non-instantaneous neutrino decoupling, gives the Standard Model prediction $N_{\nu}^{\mathrm{SM}}\simeq 3.044$~\cite{Mangano:2005cc, Fields:2019pfx, deSalas:2016ztq, Bennett:2019ewm, Akita:2020szl, Froustey:2020mcq, Bennett:2020zkv}. A representative CMB+BAO constraint from {\it Planck} 2018 gives
\begin{equation}
    N_{\mathrm{eff}} = 2.99 \pm 0.17 \quad ({\it Planck}\;2018+{\rm BAO}),
\end{equation}
which translates to approximately $\Delta N_{\mathrm{eff}}\lesssim 0.3$ and already excludes many scenarios in which axions remain coupled until well after the QCD epoch~\cite{Planck:2018vyg}. Next-generation CMB experiments (ground and space) target uncertainties of order $\sigma(N_{\mathrm{eff}})\simeq 0.03$--$0.04$, sufficient to probe the minimal thermal relic contribution $\Delta N_{\mathrm{eff}}\sim 0.03$ from pre-QCD decoupling and broad classes of ALP dark radiation models~\cite{Abazajian:2019eic, LiteBIRD:2022cnt, SimonsObservatory:2018koc}.

In summary, both thermal and non-thermal production channels make axions well-motivated DR candidates. The QCD axion provides a predictive relationship between mass and couplings (and hence $\Delta N_{\mathrm{eff}}$), while ALPs admit greater freedom and can produce observable DR without necessarily violating current stellar or laboratory bounds. Future ground and space missions will be decisive in testing many of these scenarios.

\section{Conclusions and Outlook}
\label{sec:conclusions}

In this proceeding, we have summarized the main cosmological roles of axions and axion-like particles, highlighting a limited set of mechanisms and open problems rather than providing a comprehensive review. Axions and axion-like particles remain among the most compelling extensions of the Standard Model, connecting high-energy theory to a wide range of cosmological and astrophysical phenomena. Despite the substantial progress made in understanding their production, dynamics, and observational signatures, several key questions remain open. The precise contribution of topological defects to the axion relic abundance is still uncertain, with numerical simulations producing divergent predictions for the spectrum and efficiency of axion radiation. Resolving these tensions is crucial for identifying the preferred QCD axion mass window and guiding the next generation of DM searches.

The multi-axion dynamics predicted in string-inspired axiverse scenarios introduce additional complexity, as interactions between multiple light fields can generate correlated signatures across dark sectors. Ultralight axions acting as quintessence provide a theoretically attractive framework for dynamical DE, yet the allowed field excursions, potential structure, and compatibility with quantum gravity constraints remain active topics of investigation. Similarly, the possibility that such fields could help alleviate cosmological tensions, such as the $H_0$ discrepancy, requires further study both theoretically and observationally.

On the experimental front, ongoing and planned searches are rapidly covering large portions of parameter space, but substantial regions, particularly for very low-mass or weakly coupled axions, remain unexplored. Continued development of complementary detection strategies, ranging from haloscopes and dielectric stacks to helioscopes, NMR-based detectors, dish antennas, and broadband radio observations, will be essential to provide comprehensive coverage. Looking forward, the combination of improved numerical simulations, coordinated experimental campaigns, and next-generation cosmological surveys promises to clarify the role of axions in the Universe and to test their potential as a coherent framework addressing the DM and DE problems.

\section*{Acknowledgments}
The author thanks Joshua Noah Benabou, Claudio Gatti, and Eoin \'O~Colg\'ain for useful comments. The author is grateful to the organizers of the XIX International Conference on Topics in Astroparticle and Underground Physics (TAUP 2025) for the opportunity to present this review talk, and to the Xplorer Symposia Organization Committee of the New Cornerstone Science Foundation and the Tsung-Dao Lee Institute for their hospitality and partial support of expenses. This work was supported in part by the National Natural Science Foundation of China (NSFC) under Grant No.\ 12350610240 ``Astrophysical Axion Laboratories,'' and by the Istituto Nazionale di Fisica Nucleare (INFN) through the Commissione Scientifica Nazionale~4 (CSN4) Iniziativa Specifica ``Quantum Universe'' (QGSKY). This publication is based upon work from the COST Actions ``COSMIC WISPers'' (CA21106) and ``Addressing observational tensions in cosmology with systematics and fundamental physics (CosmoVerse)'' (CA21136), both supported by COST (European Cooperation in Science and Technology).

\bibliographystyle{JHEP}
\bibliography{references.bib}

@article{SDSS:2005xqv,
    author = "Eisenstein, Daniel J. and others",
    collaboration = "SDSS",
    title = "{Detection of the Baryon Acoustic Peak in the Large-Scale Correlation Function of SDSS Luminous Red Galaxies}",
    eprint = "astro-ph/0501171",
    archivePrefix = "arXiv",
    reportNumber = "FERMILAB-PUB-05-057-A-CD",
    doi = "10.1086/466512",
    journal = "Astrophys. J.",
    volume = "633",
    pages = "560--574",
    year = "2005"
}

@article{Barbieri:2026ewj,
    author = "Barbieri, Nicola and Caloni, Luca and Gerbino, Martina and Lattanzi, Massimiliano and Visinelli, Luca",
    title = "{Beyond thermal approximations: Precise cosmological bounds on Axion-Like Particles}",
    eprint = "2602.11100",
    archivePrefix = "arXiv",
    primaryClass = "astro-ph.CO",
    reportNumber = "CA21106",
    month = "2",
    year = "2026"
}

@article{Benabou:2023npn,
    author = "Benabou, Joshua N. and Bonnefoy, Quentin and Buschmann, Malte and Kumar, Soubhik and Safdi, Benjamin R.",
    title = "{Cosmological dynamics of string theory axion strings}",
    eprint = "2312.08425",
    archivePrefix = "arXiv",
    primaryClass = "hep-ph",
    doi = "10.1103/PhysRevD.110.035021",
    journal = "Phys. Rev. D",
    volume = "110",
    number = "3",
    pages = "035021",
    year = "2024"
}

@article{Halliday:2024lca,
    author = "Halliday, Jack W. D. and others",
    title = "{Bounds on Heavy Axions with an X-Ray Free Electron Laser}",
    eprint = "2404.17333",
    archivePrefix = "arXiv",
    primaryClass = "hep-ph",
    doi = "10.1103/PhysRevLett.134.055001",
    journal = "Phys. Rev. Lett.",
    volume = "134",
    number = "5",
    pages = "055001",
    year = "2025"
}

@article{Beyer:2022ywc,
    author = "Beyer, Konstantin A. and Sarkar, Subir",
    title = "{Ruling out light axions: The writing is on the wall}",
    eprint = "2211.14635",
    archivePrefix = "arXiv",
    primaryClass = "hep-ph",
    doi = "10.21468/SciPostPhys.15.1.003",
    journal = "SciPost Phys.",
    volume = "15",
    number = "1",
    pages = "003",
    year = "2023"
}

@article{Caputo:2024oqc,
    author = "Caputo, Andrea and Raffelt, Georg",
    title = "{Astrophysical Axion Bounds: The 2024 Edition}",
    eprint = "2401.13728",
    archivePrefix = "arXiv",
    primaryClass = "hep-ph",
    reportNumber = "MPP-2024-13, CERN-TH-2024-013",
    doi = "10.22323/1.454.0041",
    journal = "PoS",
    volume = "COSMICWISPers",
    pages = "041",
    year = "2024"
}

@article{Benabou:2025kgx,
    author = "Benabou, Joshua N. and Fraser, Katherine and Reig, Mario and Safdi, Benjamin R.",
    title = "{String theory and grand unification suggest a submicroelectronvolt QCD axion}",
    eprint = "2505.15884",
    archivePrefix = "arXiv",
    primaryClass = "hep-ph",
    doi = "10.1103/lthr-97lm",
    journal = "Phys. Rev. D",
    volume = "112",
    number = "6",
    pages = "066003",
    year = "2025"
}

@article{Georgi:1986df,
    author = "Georgi, Howard and Kaplan, David B. and Randall, Lisa",
    title = "{Manifesting the Invisible Axion at Low-energies}",
    reportNumber = "HUTP-86/A004",
    doi = "10.1016/0370-2693(86)90688-X",
    journal = "Phys. Lett. B",
    volume = "169",
    pages = "73--78",
    year = "1986"
}

@book{Raffelt:1996wa,
  author    = {Raffelt, Georg G.},
  title     = {Stars as Laboratories for Fundamental Physics},
  publisher = {University of Chicago Press},
  address   = {Chicago, USA},
  year      = {1996},
  isbn      = {978-0-226-70272-8}
}

@article{VanBibber:1987rq,
    author = "Van Bibber, K. and Dagdeviren, N. R. and Koonin, S. E. and Kerman, A. and Nelson, H. N.",
    title = "{Proposed experiment to produce and detect light pseudoscalars}",
    reportNumber = "SLAC-PUB-4322",
    doi = "10.1103/PhysRevLett.59.759",
    journal = "Phys. Rev. Lett.",
    volume = "59",
    pages = "759--762",
    year = "1987"
}

@article{Veltmaat:2018dfz,
    author = "Veltmaat, Jan and Niemeyer, Jens C. and Schwabe, Bodo",
    title = "{Formation and structure of ultralight bosonic dark matter halos}",
    eprint = "1804.09647",
    archivePrefix = "arXiv",
    primaryClass = "astro-ph.CO",
    doi = "10.1103/PhysRevD.98.043509",
    journal = "Phys. Rev. D",
    volume = "98",
    number = "4",
    pages = "043509",
    year = "2018"
}

@article{Lazare:2024uvj,
    author = "Lazare, Hovav and Flitter, Jordan and Kovetz, Ely D.",
    title = "{Constraints on the fuzzy dark matter mass window from high-redshift observables}",
    eprint = "2407.19549",
    archivePrefix = "arXiv",
    primaryClass = "astro-ph.CO",
    doi = "10.1103/PhysRevD.110.123532",
    journal = "Phys. Rev. D",
    volume = "110",
    number = "12",
    pages = "123532",
    year = "2024"
}

@article{Eggemeier:2020zeg,
    author = "Eggemeier, Benedikt and Niemeyer, Jens C. and Easther, Richard",
    title = "{Formation of inflaton halos after inflation}",
    eprint = "2011.13333",
    archivePrefix = "arXiv",
    primaryClass = "astro-ph.CO",
    doi = "10.1103/PhysRevD.103.063525",
    journal = "Phys. Rev. D",
    volume = "103",
    number = "6",
    pages = "063525",
    year = "2021"
}

@article{Sipple:2024svt,
    author = "Sipple, Jackson and Lidz, Adam and Grin, Daniel and Sun, Guochao",
    title = "{Fuzzy dark matter constraints from the Hubble Frontier Fields}",
    eprint = "2407.17059",
    archivePrefix = "arXiv",
    primaryClass = "astro-ph.CO",
    doi = "10.1093/mnras/staf340",
    journal = "Mon. Not. Roy. Astron. Soc.",
    volume = "538",
    number = "3",
    pages = "1830--1842",
    year = "2025"
}

@article{Winch:2024mrt,
    author = "Winch, Harrison and Rogers, Keir K. and Hlo{\v{z}}ek, Ren{\'e}e and Marsh, David J. E.",
    title = "{High-redshift, Small-scale Tests of Ultralight Axion Dark Matter Using Hubble and Webb Galaxy UV Luminosities}",
    eprint = "2404.11071",
    archivePrefix = "arXiv",
    primaryClass = "astro-ph.CO",
    doi = "10.3847/1538-4357/ad7a73",
    journal = "Astrophys. J.",
    volume = "976",
    number = "1",
    pages = "40",
    year = "2024"
}

@article{Bird:2023pkr,
    author = "Bird, Simeon and Chang, Chia-Feng and Cui, Yanou and Yang, Daneng",
    title = "{Enhanced early galaxy formation in JWST from axion dark matter?}",
    eprint = "2307.10302",
    archivePrefix = "arXiv",
    primaryClass = "hep-ph",
    doi = "10.1016/j.physletb.2024.139062",
    journal = "Phys. Lett. B",
    volume = "858",
    pages = "139062",
    year = "2024"
}

@article{Dimopoulos:2005ac,
    author = "Dimopoulos, S. and Kachru, S. and McGreevy, J. and Wacker, Jay G.",
    title = "{N-flation}",
    eprint = "hep-th/0507205",
    archivePrefix = "arXiv",
    reportNumber = "SLAC-PUB-11016, SU-ITP-05-08",
    doi = "10.1088/1475-7516/2008/08/003",
    journal = "JCAP",
    volume = "08",
    pages = "003",
    year = "2008"
}

@article{Kim:2004rp,
    author = "Kim, Jihn E. and Nilles, Hans Peter and Peloso, Marco",
    title = "{Completing natural inflation}",
    eprint = "hep-ph/0409138",
    archivePrefix = "arXiv",
    doi = "10.1088/1475-7516/2005/01/005",
    journal = "JCAP",
    volume = "01",
    pages = "005",
    year = "2005"
}

@article{GrillidiCortona:2015jxo,
    author = "Grilli di Cortona, Giovanni and Hardy, Edward and Pardo Vega, Javier and Villadoro, Giovanni",
    title = "{The QCD axion, precisely}",
    eprint = "1511.02867",
    archivePrefix = "arXiv",
    primaryClass = "hep-ph",
    doi = "10.1007/JHEP01(2016)034",
    journal = "JHEP",
    volume = "01",
    pages = "034",
    year = "2016"
}

@article{Schive:2014hza,
    author = "Schive, Hsi-Yu and Liao, Ming-Hsuan and Woo, Tak-Pong and Wong, Shing-Kwong and Chiueh, Tzihong and Broadhurst, Tom and Hwang, W. -Y. Pauchy",
    title = "{Understanding the Core-Halo Relation of Quantum Wave Dark Matter from 3D Simulations}",
    eprint = "1407.7762",
    archivePrefix = "arXiv",
    primaryClass = "astro-ph.GA",
    doi = "10.1103/PhysRevLett.113.261302",
    journal = "Phys. Rev. Lett.",
    volume = "113",
    number = "26",
    pages = "261302",
    year = "2014"
}

@article{Schive:2014dra,
    author = "Schive, Hsi-Yu and Chiueh, Tzihong and Broadhurst, Tom",
    title = "{Cosmic Structure as the Quantum Interference of a Coherent Dark Wave}",
    eprint = "1406.6586",
    archivePrefix = "arXiv",
    primaryClass = "astro-ph.GA",
    doi = "10.1038/nphys2996",
    journal = "Nature Phys.",
    volume = "10",
    pages = "496--499",
    year = "2014"
}

@article{Zantedeschi:2026iql,
    author = "Zantedeschi, Michael",
    title = "{On the predictivity of axion dark matter in the presence of Peccei-Quinn breaking}",
    eprint = "2603.28098",
    archivePrefix = "arXiv",
    primaryClass = "hep-ph",
    month = "3",
    year = "2026"
}

@article{Figliolia:2025dtw,
    author = "Figliolia, Marco and Grippa, Francesco and Lambiase, Gaetano and Visinelli, Luca",
    title = "{Gravitational signatures of axion dark matter via parity-violating interactions}",
    eprint = "2509.12038",
    archivePrefix = "arXiv",
    primaryClass = "astro-ph.CO",
    reportNumber = "CA21106; CA21136",
    doi = "10.1103/6kqd-vntp",
    journal = "Phys. Rev. D",
    volume = "113",
    number = "2",
    pages = "023012",
    year = "2026"
}

@article{Arza:2026rsl,
    author = "Arza, A. and others",
    title = "{The COSMIC WISPers White Paper: The physics case for Weakly Interacting Slim Particles}",
    eprint = "2603.03433",
    archivePrefix = "arXiv",
    primaryClass = "hep-ph",
    reportNumber = "BARI-TH/784-26, CERN-TH-2026-016, IPPP/26/13, IFT-UAM/CSIC-26-13, KCL-PH-TH/2026-04, KEK-Cosmo-0411, KEK-TH-2804, LAPTH-008/26, MPP-2026-21, RESCEU-5/26, SLAC-PUB-260219, ST/T006994/1, ST/Y004531/1",
    month = "3",
    year = "2026"
}

@article{Bhura:2026bpy,
    author = "Bhura, Utkarsh and others",
    title = "{Axion search with telescope for radio astronomy (ASTRA): forecast for observations between 0.5 and 4{\textasciitilde}GHz}",
    eprint = "2603.13194",
    archivePrefix = "arXiv",
    primaryClass = "hep-ph",
    month = "3",
    year = "2026"
}

@article{QUAX:2025wtb,
    author = "Infirri, Giosu{\`e} Sardo and others",
    collaboration = "QUAX",
    title = "{Search for Postinflationary QCD Axions with a Quantum-Limited Tunable Microwave Receiver}",
    eprint = "2506.11589",
    archivePrefix = "arXiv",
    primaryClass = "hep-ex",
    reportNumber = "FERMILAB-PUB-25-1020-SQMS-V",
    doi = "10.1103/4dv9-72t5",
    journal = "Phys. Rev. Lett.",
    volume = "135",
    number = "21",
    pages = "211002",
    year = "2025"
}

@article{IAXO:2019mpb,
    author = "Armengaud, E. and others",
    collaboration = "IAXO",
    title = "{Physics potential of the International Axion Observatory (IAXO)}",
    eprint = "1904.09155",
    archivePrefix = "arXiv",
    primaryClass = "hep-ph",
    doi = "10.1088/1475-7516/2019/06/047",
    journal = "JCAP",
    volume = "06",
    pages = "047",
    year = "2019"
}

@article{Irastorza:2011gs,
    author = "Irastorza, I. G. and others",
    title = "{Towards a new generation axion helioscope}",
    eprint = "1103.5334",
    archivePrefix = "arXiv",
    primaryClass = "hep-ex",
    doi = "10.1088/1475-7516/2011/06/013",
    journal = "JCAP",
    volume = "06",
    pages = "013",
    year = "2011"
}

@article{Ehret:2010mh,
    author = "Ehret, Klaus and others",
    title = "{New ALPS Results on Hidden-Sector Lightweights}",
    eprint = "1004.1313",
    archivePrefix = "arXiv",
    primaryClass = "hep-ex",
    reportNumber = "DESY-10-030, MPP-2010-27",
    doi = "10.1016/j.physletb.2010.04.066",
    journal = "Phys. Lett. B",
    volume = "689",
    pages = "149--155",
    year = "2010"
}

@article{Langhoff:2022bij,
    author = "Langhoff, Kevin and Outmezguine, Nadav Joseph and Rodd, Nicholas L.",
    title = "{Irreducible Axion Background}",
    eprint = "2209.06216",
    archivePrefix = "arXiv",
    primaryClass = "hep-ph",
    reportNumber = "CERN-TH-2022-148",
    doi = "10.1103/PhysRevLett.129.241101",
    journal = "Phys. Rev. Lett.",
    volume = "129",
    number = "24",
    pages = "241101",
    year = "2022"
}

@article{CAST:2007jps,
    author = "Andriamonje, S. and others",
    collaboration = "CAST",
    title = "{An Improved limit on the axion-photon coupling from the CAST experiment}",
    eprint = "hep-ex/0702006",
    archivePrefix = "arXiv",
    doi = "10.1088/1475-7516/2007/04/010",
    journal = "JCAP",
    volume = "04",
    pages = "010",
    year = "2007"
}

@article{CAST:2017uph,
    author = "Anastassopoulos, V. and others",
    collaboration = "CAST",
    title = "{New CAST Limit on the Axion-Photon Interaction}",
    eprint = "1705.02290",
    archivePrefix = "arXiv",
    primaryClass = "hep-ex",
    doi = "10.1038/nphys4109",
    journal = "Nature Phys.",
    volume = "13",
    pages = "584--590",
    year = "2017"
}

@article{Sigl:2017sew,
    author = "Sigl, Guenter",
    title = "{Astrophysical Haloscopes}",
    eprint = "1708.08908",
    archivePrefix = "arXiv",
    primaryClass = "astro-ph.HE",
    doi = "10.1103/PhysRevD.96.103014",
    journal = "Phys. Rev. D",
    volume = "96",
    number = "10",
    pages = "103014",
    year = "2017"
}

@article{Horns:2012jf,
    author = "Horns, Dieter and Jaeckel, Joerg and Lindner, Axel and Lobanov, Andrei and Redondo, Javier and Ringwald, Andreas",
    title = "{Searching for WISPy Cold Dark Matter with a Dish Antenna}",
    eprint = "1212.2970",
    archivePrefix = "arXiv",
    primaryClass = "hep-ph",
    reportNumber = "DESY-12-227, MPP-2012-158",
    doi = "10.1088/1475-7516/2013/04/016",
    journal = "JCAP",
    volume = "04",
    pages = "016",
    year = "2013"
}

@article{Bahre:2013ywa,
    author = {B{\"a}hre, Robin and others},
    title = "{Any light particle search II {\textemdash}Technical Design Report}",
    eprint = "1302.5647",
    archivePrefix = "arXiv",
    primaryClass = "physics.ins-det",
    reportNumber = "DESY-13-030",
    doi = "10.1088/1748-0221/8/09/T09001",
    journal = "JINST",
    volume = "8",
    pages = "T09001",
    year = "2013"
}

@article{DellaValle:2014xoa,
    author = "Della Valle, F. and Milotti, E. and Ejlli, A. and Messineo, G. and Piemontese, L. and Zavattini, G. and Gastaldi, U. and Pengo, R. and Ruoso, G.",
    title = "{First results from the new PVLAS apparatus: A new limit on vacuum magnetic birefringence}",
    eprint = "1406.6518",
    archivePrefix = "arXiv",
    primaryClass = "quant-ph",
    doi = "10.1103/PhysRevD.90.092003",
    journal = "Phys. Rev. D",
    volume = "90",
    number = "9",
    pages = "092003",
    year = "2014"
}

@article{QUAX:2024fut,
    author = "Rettaroli, A. and others",
    collaboration = "QUAX",
    title = "{Search for axion dark matter with the QUAX{\textendash}LNF tunable haloscope}",
    eprint = "2402.19063",
    archivePrefix = "arXiv",
    primaryClass = "physics.ins-det",
    reportNumber = "FERMILAB-PUB-24-0511-SQMS-V",
    doi = "10.1103/PhysRevD.110.022008",
    journal = "Phys. Rev. D",
    volume = "110",
    number = "2",
    pages = "022008",
    year = "2024"
}

@article{Barbieri:2016vwg,
    author = "Barbieri, R. and Braggio, C. and Carugno, G. and Gallo, C. S. and Lombardi, A. and Ortolan, A. and Pengo, R. and Ruoso, G. and Speake, C. C.",
    title = "{Searching for galactic axions through magnetized media: the QUAX proposal}",
    eprint = "1606.02201",
    archivePrefix = "arXiv",
    primaryClass = "hep-ph",
    doi = "10.1016/j.dark.2017.01.003",
    journal = "Phys. Dark Univ.",
    volume = "15",
    pages = "135--141",
    year = "2017"
}

@article{Crescini:2018qrz,
    author = "Crescini, N. and others",
    title = "{Operation of a ferromagnetic axion haloscope at $m_a=58\,\mu$eV}",
    eprint = "1806.00310",
    archivePrefix = "arXiv",
    primaryClass = "hep-ex",
    doi = "10.1140/epjc/s10052-018-6163-8",
    journal = "Eur. Phys. J. C",
    volume = "78",
    number = "9",
    pages = "703",
    year = "2018",
    note = "[Erratum: Eur.Phys.J.C 78, 813 (2018)]"
}

@article{Alesini:2019nzq,
    author = "Alesini, D. and others",
    title = "{KLASH Conceptual Design Report}",
    eprint = "1911.02427",
    archivePrefix = "arXiv",
    primaryClass = "physics.ins-det",
    reportNumber = "INFN-19-18/LNF",
    month = "11",
    year = "2019"
}

@article{Alesini:2023qed,
    author = "Alesini, David and others",
    title = "{The future search for low-frequency axions and new physics with the FLASH resonant cavity experiment at Frascati National Laboratories}",
    eprint = "2309.00351",
    archivePrefix = "arXiv",
    primaryClass = "physics.ins-det",
    reportNumber = "CA21106; CA21136",
    doi = "10.1016/j.dark.2023.101370",
    journal = "Phys. Dark Univ.",
    volume = "42",
    pages = "101370",
    year = "2023"
}

@article{Ruoso:1992nx,
    author = "Ruoso, G. and others",
    title = "{Limits on light scalar and pseudoscalar particles from a photon regeneration experiment}",
    reportNumber = "UR-1248, ER-13065-697, FERMILAB-PUB-92-048-E",
    doi = "10.1007/BF01474722",
    journal = "Z. Phys. C",
    volume = "56",
    pages = "505--508",
    year = "1992"
}

@article{Hagmann:1990tj,
    author = "Hagmann, C. and Sikivie, P. and Sullivan, N. S. and Tanner, D. B.",
    title = "{Results from a search for cosmic axions}",
    reportNumber = "PRINT-90-0420 (FLORIDA)",
    doi = "10.1103/PhysRevD.42.1297",
    journal = "Phys. Rev. D",
    volume = "42",
    pages = "1297--1300",
    year = "1990"
}

@article{Sikivie:1985yu,
    author = "Sikivie, Pierre",
    title = "{Detection Rates for 'Invisible' Axion Searches}",
    reportNumber = "UFTP-85-5-REV, UFTP-85-5",
    doi = "10.1103/PhysRevD.36.974",
    journal = "Phys. Rev. D",
    volume = "32",
    pages = "2988",
    year = "1985",
    note = "[Erratum: Phys.Rev.D 36, 974 (1987)]"
}

@article{Burger:2018fvb,
    author = "Burger, Florian and Ilgenfritz, Ernst-Michael and Lombardo, Maria Paola and Trunin, Anton",
    title = "{Chiral observables and topology in hot QCD with two families of quarks}",
    eprint = "1805.06001",
    archivePrefix = "arXiv",
    primaryClass = "hep-lat",
    reportNumber = "HU-EP-14-64, SFB-CPP-14-103",
    doi = "10.1103/PhysRevD.98.094501",
    journal = "Phys. Rev. D",
    volume = "98",
    number = "9",
    pages = "094501",
    year = "2018"
}

@article{Lombardo:2020bvn,
    author = "Lombardo, Maria Paola and Trunin, Anton",
    title = "{Topology and axions in QCD}",
    eprint = "2005.06547",
    archivePrefix = "arXiv",
    primaryClass = "hep-lat",
    doi = "10.1142/S0217751X20300100",
    journal = "Int. J. Mod. Phys. A",
    volume = "35",
    number = "20",
    pages = "2030010",
    year = "2020"
}

@article{Kotov:2021rah,
    author = "Kotov, Andrey Yu. and Lombardo, Maria Paola and Trunin, Anton",
    title = "{QCD transition at the physical point, and its scaling window from twisted mass Wilson fermions}",
    eprint = "2105.09842",
    archivePrefix = "arXiv",
    primaryClass = "hep-lat",
    doi = "10.1016/j.physletb.2021.136749",
    journal = "Phys. Lett. B",
    volume = "823",
    pages = "136749",
    year = "2021"
}

@article{Borsanyi:2022xml,
    author = "Borsanyi, S. and R., Kara and Fodor, Z. and Godzieba, D. A. and Parotto, P. and Sexty, D.",
    title = "{Precision study of the continuum SU(3) Yang-Mills theory: How to use parallel tempering to improve on supercritical slowing down for first order phase transitions}",
    eprint = "2202.05234",
    archivePrefix = "arXiv",
    primaryClass = "hep-lat",
    doi = "10.1103/PhysRevD.105.074513",
    journal = "Phys. Rev. D",
    volume = "105",
    number = "7",
    pages = "074513",
    year = "2022"
}

@article{Planck:2018vyg,
    author = "Aghanim, N. and others",
    collaboration = "Planck",
    title = "{Planck 2018 results. VI. Cosmological parameters}",
    eprint = "1807.06209",
    archivePrefix = "arXiv",
    primaryClass = "astro-ph.CO",
    doi = "10.1051/0004-6361/201833910",
    journal = "Astron. Astrophys.",
    volume = "641",
    pages = "A6",
    year = "2020",
    note = "[Erratum: Astron.Astrophys. 652, C4 (2021)]"
}

@article{Planck:2019nip,
    author = "Aghanim, N. and others",
    collaboration = "Planck",
    title = "{Planck 2018 results. V. CMB power spectra and likelihoods}",
    eprint = "1907.12875",
    archivePrefix = "arXiv",
    primaryClass = "astro-ph.CO",
    doi = "10.1051/0004-6361/201936386",
    journal = "Astron. Astrophys.",
    volume = "641",
    pages = "A5",
    year = "2020"
}

@article{Peccei:1977hh,
    author = "Peccei, R. D. and Quinn, Helen R.",
    title = "{CP Conservation in the Presence of Instantons}",
    reportNumber = "ITP-568-STANFORD",
    doi = "10.1103/PhysRevLett.38.1440",
    journal = "Phys. Rev. Lett.",
    volume = "38",
    pages = "1440--1443",
    year = "1977"
}

@article{Weinberg:1977ma,
    author = "Weinberg, Steven",
    title = "{A New Light Boson?}",
    reportNumber = "HUTP-77/A074",
    doi = "10.1103/PhysRevLett.40.223",
    journal = "Phys. Rev. Lett.",
    volume = "40",
    pages = "223--226",
    year = "1978"
}

@article{Wilczek:1977pj,
    author = "Wilczek, Frank",
    title = "{Problem of Strong  $P$  and  $T$  Invariance in the Presence of Instantons}",
    reportNumber = "Print-77-0939 (COLUMBIA)",
    doi = "10.1103/PhysRevLett.40.279",
    journal = "Phys. Rev. Lett.",
    volume = "40",
    pages = "279--282",
    year = "1978"
}

@article{Preskill:1982cy,
    author = "Preskill, John and Wise, Mark B. and Wilczek, Frank",
    editor = "Srednicki, M. A.",
    title = "{Cosmology of the Invisible Axion}",
    reportNumber = "HUTP-82-A048, NSF-ITP-82-103",
    doi = "10.1016/0370-2693(83)90637-8",
    journal = "Phys. Lett. B",
    volume = "120",
    pages = "127--132",
    year = "1983"
}

@article{Abbott:1982af,
    author = "Abbott, L. F. and Sikivie, P.",
    editor = "Srednicki, M. A.",
    title = "{A Cosmological Bound on the Invisible Axion}",
    reportNumber = "PRINT-82-0695 (BRANDEIS)",
    doi = "10.1016/0370-2693(83)90638-X",
    journal = "Phys. Lett. B",
    volume = "120",
    pages = "133--136",
    year = "1983"
}

@article{Dine:1982ah,
    author = "Dine, Michael and Fischler, Willy",
    editor = "Srednicki, M. A.",
    title = "{The Not So Harmless Axion}",
    reportNumber = "UPR-0201T",
    doi = "10.1016/0370-2693(83)90639-1",
    journal = "Phys. Lett. B",
    volume = "120",
    pages = "137--141",
    year = "1983"
}

@article{Kibble:1976sj,
    author = "Kibble, T. W. B.",
    title = "{Topology of Cosmic Domains and Strings}",
    reportNumber = "ICTP/75/5",
    doi = "10.1088/0305-4470/9/8/029",
    journal = "J. Phys. A",
    volume = "9",
    pages = "1387--1398",
    year = "1976"
}

@article{Vilenkin:1982ks,
    author = "Vilenkin, A. and Everett, A. E.",
    title = "{Cosmic Strings and Domain Walls in Models with Goldstone and PseudoGoldstone Bosons}",
    doi = "10.1103/PhysRevLett.48.1867",
    journal = "Phys. Rev. Lett.",
    volume = "48",
    pages = "1867--1870",
    year = "1982"
}

@article{Lyth:1992tw,
    author = "Lyth, David H. and Stewart, Ewan D.",
    title = "{Axions and inflation: String formation during inflation}",
    reportNumber = "LANC-TH-92-03",
    doi = "10.1103/PhysRevD.46.532",
    journal = "Phys. Rev. D",
    volume = "46",
    pages = "532--538",
    year = "1992"
}

@article{Davis:1986xc,
    author = "Davis, Richard Lynn",
    title = "{Cosmic Axions from Cosmic Strings}",
    reportNumber = "SLAC-PUB-3895",
    doi = "10.1016/0370-2693(86)90300-X",
    journal = "Phys. Lett. B",
    volume = "180",
    pages = "225--230",
    year = "1986"
}

@article{Harari:1987ht,
    author = "Harari, Diego and Sikivie, P.",
    title = "{On the Evolution of Global Strings in the Early Universe}",
    reportNumber = "UFTP-87-4",
    doi = "10.1016/0370-2693(87)90032-3",
    journal = "Phys. Lett. B",
    volume = "195",
    pages = "361--365",
    year = "1987"
}

@article{Davis:1989nj,
    author = "Davis, R. L. and Shellard, E. P. S.",
    title = "{DO AXIONS NEED INFLATION?}",
    reportNumber = "NSF-ITP-88-195, MIT-CTP-1685, CSR-AT-89-07",
    doi = "10.1016/0550-3213(89)90187-9",
    journal = "Nucl. Phys. B",
    volume = "324",
    pages = "167--186",
    year = "1989"
}

@article{Battye:1994au,
    author = "Battye, R. A. and Shellard, E. P. S.",
    title = "{Axion string constraints}",
    eprint = "astro-ph/9403018",
    archivePrefix = "arXiv",
    reportNumber = "DAMTP-R-94-8",
    doi = "10.1103/PhysRevLett.73.2954",
    journal = "Phys. Rev. Lett.",
    volume = "73",
    pages = "2954--2957",
    year = "1994",
    note = "[Erratum: Phys.Rev.Lett. 76, 2203--2204 (1996)]"
}

@article{Chang:1998tb,
    author = "Chang, Sanghyeon and Hagmann, C. and Sikivie, P.",
    title = "{Studies of the motion and decay of axion walls bounded by strings}",
    eprint = "hep-ph/9807374",
    archivePrefix = "arXiv",
    reportNumber = "UFIFT-HEP-98-12",
    doi = "10.1103/PhysRevD.59.023505",
    journal = "Phys. Rev. D",
    volume = "59",
    pages = "023505",
    year = "1999"
}

@article{Hiramatsu:2010yn,
    author = "Hiramatsu, Takashi and Kawasaki, Masahiro and Saikawa, Ken'ichi",
    title = "{Evolution of String-Wall Networks and Axionic Domain Wall Problem}",
    eprint = "1012.4558",
    archivePrefix = "arXiv",
    primaryClass = "astro-ph.CO",
    reportNumber = "ICRR-REPORT-577-2010-10, IPMU10-0221, YITP-10-110",
    doi = "10.1088/1475-7516/2011/08/030",
    journal = "JCAP",
    volume = "08",
    pages = "030",
    year = "2011"
}

@article{Hiramatsu:2012gg,
    author = "Hiramatsu, Takashi and Kawasaki, Masahiro and Saikawa, Ken'ichi and Sekiguchi, Toyokazu",
    title = "{Production of dark matter axions from collapse of string-wall systems}",
    eprint = "1202.5851",
    archivePrefix = "arXiv",
    primaryClass = "hep-ph",
    reportNumber = "ICRR-REPORT-608-2011-25, IPMU12-0025, YITP-12-9",
    doi = "10.1103/PhysRevD.85.105020",
    journal = "Phys. Rev. D",
    volume = "85",
    pages = "105020",
    year = "2012",
    note = "[Erratum: Phys.Rev.D 86, 089902 (2012)]"
}

@article{Kawasaki:2018bzv,
    author = "Kawasaki, Masahiro and Sekiguchi, Toyokazu and Yamaguchi, Masahide and Yokoyama, Jun'ichi",
    title = "{Long-term dynamics of cosmological axion strings}",
    eprint = "1806.05566",
    archivePrefix = "arXiv",
    primaryClass = "hep-ph",
    reportNumber = "RESCEU-8/18, RESCEU-8-18",
    doi = "10.1093/ptep/pty098",
    journal = "PTEP",
    volume = "2018",
    number = "9",
    pages = "091E01",
    year = "2018"
}

@article{Saikawa:2012uk,
    author = "Saikawa, Ken'ichi and Yamaguchi, Masahide",
    title = "{Evolution and thermalization of dark matter axions in the condensed regime}",
    eprint = "1210.7080",
    archivePrefix = "arXiv",
    primaryClass = "hep-ph",
    reportNumber = "ICRR-REPORT-632-2012-21",
    doi = "10.1103/PhysRevD.87.085010",
    journal = "Phys. Rev. D",
    volume = "87",
    number = "8",
    pages = "085010",
    year = "2013"
}

@article{Vaquero:2018tib,
    author = "Vaquero, Alejandro and Redondo, Javier and Stadler, Julia",
    title = "{Early seeds of axion miniclusters}",
    eprint = "1809.09241",
    archivePrefix = "arXiv",
    primaryClass = "astro-ph.CO",
    doi = "10.1088/1475-7516/2019/04/012",
    journal = "JCAP",
    volume = "04",
    pages = "012",
    year = "2019"
}

@article{Kawasaki:2014sqa,
    author = "Kawasaki, Masahiro and Saikawa, Ken'ichi and Sekiguchi, Toyokazu",
    title = "{Axion dark matter from topological defects}",
    eprint = "1412.0789",
    archivePrefix = "arXiv",
    primaryClass = "hep-ph",
    reportNumber = "ICRR-REPORT-696-2014-22, IPMU14-0348",
    doi = "10.1103/PhysRevD.91.065014",
    journal = "Phys. Rev. D",
    volume = "91",
    number = "6",
    pages = "065014",
    year = "2015"
}

@article{Ringwald:2015dsf,
    author = "Ringwald, Andreas and Saikawa, Ken'ichi",
    title = "{Axion dark matter in the post-inflationary Peccei-Quinn symmetry breaking scenario}",
    eprint = "1512.06436",
    archivePrefix = "arXiv",
    primaryClass = "hep-ph",
    reportNumber = "DESY-15-250",
    doi = "10.1103/PhysRevD.93.085031",
    journal = "Phys. Rev. D",
    volume = "93",
    number = "8",
    pages = "085031",
    year = "2016",
    note = "[Addendum: Phys.Rev.D 94, 049908 (2016)]"
}

@article{Harigaya:2018ooc,
    author = "Harigaya, Keisuke and Kawasaki, Masahiro",
    title = "{QCD axion dark matter from long-lived domain walls during matter domination}",
    eprint = "1802.00579",
    archivePrefix = "arXiv",
    primaryClass = "hep-ph",
    reportNumber = "IPMU-18-0027",
    doi = "10.1016/j.physletb.2018.04.056",
    journal = "Phys. Lett. B",
    volume = "782",
    pages = "1--5",
    year = "2018"
}

@article{Caputo:2019wsd,
    author = "Caputo, Andrea and Reig, Mario",
    title = "{Cosmic implications of a low-scale solution to the axion domain wall problem}",
    eprint = "1905.13116",
    archivePrefix = "arXiv",
    primaryClass = "hep-ph",
    reportNumber = "IFIC/19-29",
    doi = "10.1103/PhysRevD.100.063530",
    journal = "Phys. Rev. D",
    volume = "100",
    number = "6",
    pages = "063530",
    year = "2019"
}

@article{Arvanitaki:2009fg,
    author = "Arvanitaki, Asimina and Dimopoulos, Savas and Dubovsky, Sergei and Kaloper, Nemanja and March-Russell, John",
    title = "{String Axiverse}",
    eprint = "0905.4720",
    archivePrefix = "arXiv",
    primaryClass = "hep-th",
    doi = "10.1103/PhysRevD.81.123530",
    journal = "Phys. Rev. D",
    volume = "81",
    pages = "123530",
    year = "2010"
}

@article{Witten:1984dg,
    author = "Witten, Edward",
    title = "{Some Properties of O(32) Superstrings}",
    reportNumber = "Print-84-0838 (PRINCETON)",
    doi = "10.1016/0370-2693(84)90422-2",
    journal = "Phys. Lett. B",
    volume = "149",
    pages = "351--356",
    year = "1984"
}

@article{Fox:2004kb,
    author = "Fox, Patrick and Pierce, Aaron and Thomas, Scott D.",
    title = "{Probing a QCD string axion with precision cosmological measurements}",
    eprint = "hep-th/0409059",
    archivePrefix = "arXiv",
    reportNumber = "SLAC-PUB-10030, SU-ITP-03-19, SCIPP-2003-04",
    month = "9",
    year = "2004"
}

@article{Conlon:2006tq,
    author = "Conlon, Joseph P.",
    title = "{The QCD axion and moduli stabilisation}",
    eprint = "hep-th/0602233",
    archivePrefix = "arXiv",
    reportNumber = "DAMTP-2006-17",
    doi = "10.1088/1126-6708/2006/05/078",
    journal = "JHEP",
    volume = "05",
    pages = "078",
    year = "2006"
}

@article{Svrcek:2006yi,
    author = "Svrcek, Peter and Witten, Edward",
    title = "{Axions In String Theory}",
    eprint = "hep-th/0605206",
    archivePrefix = "arXiv",
    reportNumber = "SLAC-PUB-11894",
    doi = "10.1088/1126-6708/2006/06/051",
    journal = "JHEP",
    volume = "06",
    pages = "051",
    year = "2006"
}

@article{Cicoli:2012sz,
    author = "Cicoli, Michele and Goodsell, Mark and Ringwald, Andreas",
    title = "{The type IIB string axiverse and its low-energy phenomenology}",
    eprint = "1206.0819",
    archivePrefix = "arXiv",
    primaryClass = "hep-th",
    reportNumber = "DESY-12-058, CERN-PH-TH-2012-153",
    doi = "10.1007/JHEP10(2012)146",
    journal = "JHEP",
    volume = "10",
    pages = "146",
    year = "2012"
}

@article{Demirtas:2018akl,
    author = "Demirtas, Mehmet and Long, Cody and McAllister, Liam and Stillman, Mike",
    title = "{The Kreuzer-Skarke Axiverse}",
    eprint = "1808.01282",
    archivePrefix = "arXiv",
    primaryClass = "hep-th",
    doi = "10.1007/JHEP04(2020)138",
    journal = "JHEP",
    volume = "04",
    pages = "138",
    year = "2020"
}

@article{Bertone:2004pz,
    author = "Bertone, Gianfranco and Hooper, Dan and Silk, Joseph",
    title = "{Particle dark matter: Evidence, candidates and constraints}",
    eprint = "hep-ph/0404175",
    archivePrefix = "arXiv",
    reportNumber = "FERMILAB-PUB-04-047-A",
    doi = "10.1016/j.physrep.2004.08.031",
    journal = "Phys. Rept.",
    volume = "405",
    pages = "279--390",
    year = "2005"
}

@article{Raffelt:1987yt,
    author = "Raffelt, Georg and Seckel, David",
    title = "{Bounds on Exotic Particle Interactions from SN 1987a}",
    reportNumber = "SCIPP-87/107",
    doi = "10.1103/PhysRevLett.60.1793",
    journal = "Phys. Rev. Lett.",
    volume = "60",
    pages = "1793",
    year = "1988"
}

@article{Cirelli:2024ssz,
    author = "Cirelli, Marco and Strumia, Alessandro and Zupan, Jure",
    title = "{Dark Matter}",
    eprint = "2406.01705",
    archivePrefix = "arXiv",
    primaryClass = "hep-ph",
    month = "6",
    year = "2024"
}

@book{Marsh:2024ury,
    author = "Marsh, David J. E. and Ellis, David and Mehta, Viraf M.",
    title = "{Dark Matter: Evidence, Theory, and Constraints}",
    doi = "10.1515/9780691249711",
    isbn = "978-0-691-24971-1, 978-0-691-24952-0",
    publisher = "Princeton University Press",
    series = "Princeton Series in Astrophysics",
    month = "10",
    year = "2024"
}

@article{Ellis:2020gtq,
    author = "Ellis, David and Marsh, David J. E. and Behrens, Christoph",
    title = "{Axion Miniclusters Made Easy}",
    eprint = "2006.08637",
    archivePrefix = "arXiv",
    primaryClass = "astro-ph.CO",
    doi = "10.1103/PhysRevD.103.083525",
    journal = "Phys. Rev. D",
    volume = "103",
    number = "8",
    pages = "083525",
    year = "2021"
}

@article{Yin:2024xov,
    author = "Yin, Ziwen and Visinelli, Luca",
    title = "{Axion star condensation around primordial black holes and microlensing limits}",
    eprint = "2404.10340",
    archivePrefix = "arXiv",
    primaryClass = "hep-ph",
    reportNumber = "CA21106; CA21136",
    doi = "10.1088/1475-7516/2024/10/013",
    journal = "JCAP",
    volume = "10",
    pages = "013",
    year = "2024"
}

@article{Schiappacasse:2017ham,
    author = "Schiappacasse, Enrico D. and Hertzberg, Mark P.",
    title = "{Analysis of Dark Matter Axion Clumps with Spherical Symmetry}",
    eprint = "1710.04729",
    archivePrefix = "arXiv",
    primaryClass = "hep-ph",
    doi = "10.1088/1475-7516/2018/01/037",
    journal = "JCAP",
    volume = "01",
    pages = "037",
    year = "2018",
    note = "[Erratum: JCAP 03, E01 (2018)]"
}

@article{Hu:2000ke,
    author = "Hu, Wayne and Barkana, Rennan and Gruzinov, Andrei",
    title = "{Cold and fuzzy dark matter}",
    eprint = "astro-ph/0003365",
    archivePrefix = "arXiv",
    doi = "10.1103/PhysRevLett.85.1158",
    journal = "Phys. Rev. Lett.",
    volume = "85",
    pages = "1158--1161",
    year = "2000"
}

@article{Hui:2016ltb,
    author = "Hui, Lam and Ostriker, Jeremiah P. and Tremaine, Scott and Witten, Edward",
    title = "{Ultralight scalars as cosmological dark matter}",
    eprint = "1610.08297",
    archivePrefix = "arXiv",
    primaryClass = "astro-ph.CO",
    doi = "10.1103/PhysRevD.95.043541",
    journal = "Phys. Rev. D",
    volume = "95",
    number = "4",
    pages = "043541",
    year = "2017"
}

@article{Marsh:2015xka,
    author = "Marsh, David J. E.",
    title = "{Axion Cosmology}",
    eprint = "1510.07633",
    archivePrefix = "arXiv",
    primaryClass = "astro-ph.CO",
    reportNumber = "KCL-PH-TH-2015-50",
    doi = "10.1016/j.physrep.2016.06.005",
    journal = "Phys. Rept.",
    volume = "643",
    pages = "1--79",
    year = "2016"
}

@article{Frieman:1995pm,
    author = "Frieman, Joshua A. and Hill, Christopher T. and Stebbins, Albert and Waga, Ioav",
    title = "{Cosmology with ultralight pseudo Nambu-Goldstone bosons}",
    eprint = "astro-ph/9505060",
    archivePrefix = "arXiv",
    reportNumber = "FERMILAB-PUB-95-066-A",
    doi = "10.1103/PhysRevLett.75.2077",
    journal = "Phys. Rev. Lett.",
    volume = "75",
    pages = "2077--2080",
    year = "1995"
}

@article{Kaloper:2005aj,
    author = "Kaloper, Nemanja and Sorbo, Lorenzo",
    title = "{Of pngb quintessence}",
    eprint = "astro-ph/0511543",
    archivePrefix = "arXiv",
    doi = "10.1088/1475-7516/2006/04/007",
    journal = "JCAP",
    volume = "04",
    pages = "007",
    year = "2006"
}

@article{Nomura:2000yk,
    author = "Nomura, Yasunori and Watari, T. and Yanagida, T.",
    title = "{Quintessence axion potential induced by electroweak instanton effects}",
    eprint = "hep-ph/0004182",
    archivePrefix = "arXiv",
    reportNumber = "UT-883",
    doi = "10.1016/S0370-2693(00)00605-5",
    journal = "Phys. Lett. B",
    volume = "484",
    pages = "103--111",
    year = "2000"
}

@article{Visinelli:2018utg,
    author = "Visinelli, Luca and Vagnozzi, Sunny",
    title = "{Cosmological window onto the string axiverse and the supersymmetry breaking scale}",
    eprint = "1809.06382",
    archivePrefix = "arXiv",
    primaryClass = "hep-ph",
    reportNumber = "NORDITA-2018-051",
    doi = "10.1103/PhysRevD.99.063517",
    journal = "Phys. Rev. D",
    volume = "99",
    number = "6",
    pages = "063517",
    year = "2019"
}

@article{Freese:1990rb,
    author = "Freese, Katherine and Frieman, Joshua A. and Olinto, Angela V.",
    title = "{Natural inflation with pseudo - Nambu-Goldstone bosons}",
    reportNumber = "FERMILAB-PUB-90-177-A",
    doi = "10.1103/PhysRevLett.65.3233",
    journal = "Phys. Rev. Lett.",
    volume = "65",
    pages = "3233--3236",
    year = "1990"
}

@article{Adams:1992bn,
    author = "Adams, Fred C. and Bond, J. Richard and Freese, Katherine and Frieman, Joshua A. and Olinto, Angela V.",
    title = "{Natural inflation: Particle physics models, power law spectra for large scale structure, and constraints from COBE}",
    eprint = "hep-ph/9207245",
    archivePrefix = "arXiv",
    reportNumber = "FERMILAB-PUB-92-202-A",
    doi = "10.1103/PhysRevD.47.426",
    journal = "Phys. Rev. D",
    volume = "47",
    pages = "426--455",
    year = "1993"
}

@article{Silverstein:2008sg,
    author = "Silverstein, Eva and Westphal, Alexander",
    title = "{Monodromy in the CMB: Gravity Waves and String Inflation}",
    eprint = "0803.3085",
    archivePrefix = "arXiv",
    primaryClass = "hep-th",
    reportNumber = "SU-ITP-08-07, SLAC-PUB-13183",
    doi = "10.1103/PhysRevD.78.106003",
    journal = "Phys. Rev. D",
    volume = "78",
    pages = "106003",
    year = "2008"
}

@article{McAllister:2008hb,
    author = "McAllister, Liam and Silverstein, Eva and Westphal, Alexander",
    title = "{Gravity Waves and Linear Inflation from Axion Monodromy}",
    eprint = "0808.0706",
    archivePrefix = "arXiv",
    primaryClass = "hep-th",
    reportNumber = "SLAC-PUB-13357, SU-ITP-08-15",
    doi = "10.1103/PhysRevD.82.046003",
    journal = "Phys. Rev. D",
    volume = "82",
    pages = "046003",
    year = "2010"
}

@article{Visinelli:2011jy,
    author = "Visinelli, Luca",
    title = "{Natural Warm Inflation}",
    eprint = "1107.3523",
    archivePrefix = "arXiv",
    primaryClass = "astro-ph.CO",
    doi = "10.1088/1475-7516/2011/09/013",
    journal = "JCAP",
    volume = "09",
    pages = "013",
    year = "2011"
}

@article{Arias:2012az,
    author = "Arias, Paola and Cadamuro, Davide and Goodsell, Mark and Jaeckel, Joerg and Redondo, Javier and Ringwald, Andreas",
    title = "{WISPy Cold Dark Matter}",
    eprint = "1201.5902",
    archivePrefix = "arXiv",
    primaryClass = "hep-ph",
    reportNumber = "DESY-11-226, MPP-2011-140, CERN-PH-TH-2011-323, IPPP-11-80, DCPT-11-160",
    doi = "10.1088/1475-7516/2012/06/013",
    journal = "JCAP",
    volume = "06",
    pages = "013",
    year = "2012"
}

@article{Graf:2010tv,
    author = "Graf, Peter and Steffen, Frank Daniel",
    title = "{Thermal axion production in the primordial quark-gluon plasma}",
    eprint = "1008.4528",
    archivePrefix = "arXiv",
    primaryClass = "hep-ph",
    reportNumber = "MPP-2010-20",
    doi = "10.1103/PhysRevD.83.075011",
    journal = "Phys. Rev. D",
    volume = "83",
    pages = "075011",
    year = "2011"
}

@article{DEramo:2018vss,
    author = "D'Eramo, Francesco and Ferreira, Ricardo Z. and Notari, Alessio and Bernal, Jos{\'e} Luis",
    title = "{Hot Axions and the $H_0$ tension}",
    eprint = "1808.07430",
    archivePrefix = "arXiv",
    primaryClass = "hep-ph",
    doi = "10.1088/1475-7516/2018/11/014",
    journal = "JCAP",
    volume = "11",
    pages = "014",
    year = "2018"
}

@article{Caloni:2022uya,
    author = "Caloni, Luca and Gerbino, Martina and Lattanzi, Massimiliano and Visinelli, Luca",
    title = "{Novel cosmological bounds on thermally-produced axion-like particles}",
    eprint = "2205.01637",
    archivePrefix = "arXiv",
    primaryClass = "astro-ph.CO",
    doi = "10.1088/1475-7516/2022/09/021",
    journal = "JCAP",
    volume = "09",
    pages = "021",
    year = "2022"
}

@article{Salvio:2013iaa,
    author = "Salvio, Alberto and Strumia, Alessandro and Xue, Wei",
    title = "{Thermal axion production}",
    eprint = "1310.6982",
    archivePrefix = "arXiv",
    primaryClass = "hep-ph",
    reportNumber = "FTUAM-13-29, IFT-UAM-CSIC-13-113",
    doi = "10.1088/1475-7516/2014/01/011",
    journal = "JCAP",
    volume = "01",
    pages = "011",
    year = "2014"
}

@article{Wantz:2009it,
    author = "Wantz, Olivier and Shellard, E. P. S.",
    title = "{Axion Cosmology Revisited}",
    eprint = "0910.1066",
    archivePrefix = "arXiv",
    primaryClass = "astro-ph.CO",
    doi = "10.1103/PhysRevD.82.123508",
    journal = "Phys. Rev. D",
    volume = "82",
    pages = "123508",
    year = "2010"
}

@article{Archidiacono:2013cha,
    author = "Archidiacono, Maria and Hannestad, Steen and Mirizzi, Alessandro and Raffelt, Georg and Wong, Yvonne Y. Y.",
    title = "{Axion hot dark matter bounds after Planck}",
    eprint = "1307.0615",
    archivePrefix = "arXiv",
    primaryClass = "astro-ph.CO",
    reportNumber = "MPP-2013-113",
    doi = "10.1088/1475-7516/2013/10/020",
    journal = "JCAP",
    volume = "10",
    pages = "020",
    year = "2013"
}

@article{Cadamuro:2011fd,
    author = "Cadamuro, Davide and Redondo, Javier",
    title = "{Cosmological bounds on pseudo Nambu-Goldstone bosons}",
    eprint = "1110.2895",
    archivePrefix = "arXiv",
    primaryClass = "hep-ph",
    reportNumber = "MPP-2011-116",
    doi = "10.1088/1475-7516/2012/02/032",
    journal = "JCAP",
    volume = "02",
    pages = "032",
    year = "2012"
}

@article{Essig:2013goa,
    author = "Essig, Rouven and Kuflik, Eric and McDermott, Samuel D. and Volansky, Tomer and Zurek, Kathryn M.",
    title = "{Constraining Light Dark Matter with Diffuse X-Ray and Gamma-Ray Observations}",
    eprint = "1309.4091",
    archivePrefix = "arXiv",
    primaryClass = "hep-ph",
    reportNumber = "YITP-SB-29-13, FERMILAB-PUB-13-377-A-T, MCTP-13-27",
    doi = "10.1007/JHEP11(2013)193",
    journal = "JHEP",
    volume = "11",
    pages = "193",
    year = "2013"
}

@article{Conlon:2013txa,
    author = "Conlon, Joseph P. and Marsh, M. C. David",
    title = "{Excess Astrophysical Photons from a 0.1{\textendash}1 keV Cosmic Axion Background}",
    eprint = "1305.3603",
    archivePrefix = "arXiv",
    primaryClass = "astro-ph.CO",
    doi = "10.1103/PhysRevLett.111.151301",
    journal = "Phys. Rev. Lett.",
    volume = "111",
    number = "15",
    pages = "151301",
    year = "2013"
}

@article{Grin:2006aw,
    author = "Grin, Daniel and Covone, Giovanni and Kneib, Jean-Paul and Kamionkowski, Marc and Blain, Andrew and Jullo, Eric",
    title = "{A Telescope Search for Decaying Relic Axions}",
    eprint = "astro-ph/0611502",
    archivePrefix = "arXiv",
    doi = "10.1103/PhysRevD.75.105018",
    journal = "Phys. Rev. D",
    volume = "75",
    pages = "105018",
    year = "2007"
}

@article{Cheng:2025cmb,
    author = "Cheng, Hanyu and Yin, Ziwen and Di Valentino, Eleonora and Marsh, David J. E. and Visinelli, Luca",
    title = "{Constraining exotic high-$z$ reionization histories with Gaussian processes and the cosmic microwave background}",
    eprint = "2506.19096",
    archivePrefix = "arXiv",
    primaryClass = "astro-ph.CO",
    reportNumber = "CA21106; CA21136",
    month = "6",
    year = "2025"
}

@article{Capozzi:2023xie,
    author = "Capozzi, Francesco and Ferreira, Ricardo Z. and Lopez-Honorez, Laura and Mena, Olga",
    title = "{CMB and Lyman-{\ensuremath{\alpha}} constraints on dark matter decays to photons}",
    eprint = "2303.07426",
    archivePrefix = "arXiv",
    primaryClass = "astro-ph.CO",
    reportNumber = "ULB-TH/23-03",
    doi = "10.1088/1475-7516/2023/06/060",
    journal = "JCAP",
    volume = "06",
    pages = "060",
    year = "2023"
}

@article{Slatyer:2024nbm,
    author = "Slatyer, Tracy R.",
    title = "{What does cosmology teach us about non-gravitational properties of dark matter?}",
    doi = "10.1016/j.nuclphysb.2024.116468",
    journal = "Nucl. Phys. B",
    volume = "1003",
    pages = "116468",
    year = "2024"
}

@article{Depta:2020zbh,
    author = "Depta, Paul Frederik and Hufnagel, Marco and Schmidt-Hoberg, Kai",
    title = "{Updated BBN constraints on electromagnetic decays of MeV-scale particles}",
    eprint = "2011.06519",
    archivePrefix = "arXiv",
    primaryClass = "hep-ph",
    reportNumber = "DESY-20-160, DESY 20-160, ULB-TH/20-15",
    doi = "10.1088/1475-7516/2021/04/011",
    journal = "JCAP",
    volume = "04",
    pages = "011",
    year = "2021"
}

@article{Balazs:2022tjl,
    author = "Bal{\'a}zs, Csaba and others",
    title = "{Cosmological constraints on decaying axion-like particles: a global analysis}",
    eprint = "2205.13549",
    archivePrefix = "arXiv",
    primaryClass = "astro-ph.CO",
    reportNumber = "gambit-physics-2022, KCL-PH-TH/2022-23, TTP22-034",
    doi = "10.1088/1475-7516/2022/12/027",
    journal = "JCAP",
    volume = "12",
    pages = "027",
    year = "2022"
}

@article{Yin:2025amn,
    author = "Yin, Ziwen and Cheng, Hanyu and Di Valentino, Eleonora and Gendler, Naomi and Marsh, David J. E. and Visinelli, Luca",
    title = "{Constraining the axiverse with reionization}",
    eprint = "2507.03535",
    archivePrefix = "arXiv",
    primaryClass = "hep-ph",
    reportNumber = "CA21106; CA21136",
    month = "7",
    year = "2025"
}

@article{Millea:2015qra,
    author = "Millea, Marius and Knox, Lloyd and Fields, Brian",
    title = "{New Bounds for Axions and Axion-Like Particles with keV-GeV Masses}",
    eprint = "1501.04097",
    archivePrefix = "arXiv",
    primaryClass = "astro-ph.CO",
    doi = "10.1103/PhysRevD.92.023010",
    journal = "Phys. Rev. D",
    volume = "92",
    number = "2",
    pages = "023010",
    year = "2015"
}

@article{Depta:2020wmr,
    author = "Depta, Paul Frederik and Hufnagel, Marco and Schmidt-Hoberg, Kai",
    title = "{Robust cosmological constraints on axion-like particles}",
    eprint = "2002.08370",
    archivePrefix = "arXiv",
    primaryClass = "hep-ph",
    reportNumber = "DESY-20-003, DESY 20-003",
    doi = "10.1088/1475-7516/2020/05/009",
    journal = "JCAP",
    volume = "05",
    pages = "009",
    year = "2020"
}

@article{Jaeckel:2015jla,
    author = "Jaeckel, Joerg and Spannowsky, Michael",
    title = "{Probing MeV to 90 GeV axion-like particles with LEP and LHC}",
    eprint = "1509.00476",
    archivePrefix = "arXiv",
    primaryClass = "hep-ph",
    doi = "10.1016/j.physletb.2015.12.037",
    journal = "Phys. Lett. B",
    volume = "753",
    pages = "482--487",
    year = "2016"
}

@article{Dolan:2017osp,
    author = "Dolan, Matthew J. and Ferber, Torben and Hearty, Christopher and Kahlhoefer, Felix and Schmidt-Hoberg, Kai",
    title = "{Revised constraints and Belle II sensitivity for visible and invisible axion-like particles}",
    eprint = "1709.00009",
    archivePrefix = "arXiv",
    primaryClass = "hep-ph",
    reportNumber = "DESY-17-127",
    doi = "10.1007/JHEP12(2017)094",
    journal = "JHEP",
    volume = "12",
    pages = "094",
    year = "2017",
    note = "[Erratum: JHEP 03, 190 (2021)]"
}

@article{Bauer:2017ris,
    author = "Bauer, Martin and Neubert, Matthias and Thamm, Andrea",
    title = "{Collider Probes of Axion-Like Particles}",
    eprint = "1708.00443",
    archivePrefix = "arXiv",
    primaryClass = "hep-ph",
    reportNumber = "MITP-17-047",
    doi = "10.1007/JHEP12(2017)044",
    journal = "JHEP",
    volume = "12",
    pages = "044",
    year = "2017"
}

@article{Banerjee:2017hhz,
    author = "Banerjee, D. and others",
    collaboration = "NA64",
    title = "{Search for vector mediator of Dark Matter production in invisible decay mode}",
    eprint = "1710.00971",
    archivePrefix = "arXiv",
    primaryClass = "hep-ex",
    doi = "10.1103/PhysRevD.97.072002",
    journal = "Phys. Rev. D",
    volume = "97",
    number = "7",
    pages = "072002",
    year = "2018"
}

@article{Bernreuther:2019pfb,
    author = {Bernreuther, Elias and Kahlhoefer, Felix and Kr{\"a}mer, Michael and Tunney, Patrick},
    title = "{Strongly interacting dark sectors in the early Universe and at the LHC through a simplified portal}",
    eprint = "1907.04346",
    archivePrefix = "arXiv",
    primaryClass = "hep-ph",
    reportNumber = "TTK-19-25, P3H-19-019",
    doi = "10.1007/JHEP01(2020)162",
    journal = "JHEP",
    volume = "01",
    pages = "162",
    year = "2020"
}

@article{Arvanitaki:2010sy,
    author = "Arvanitaki, Asimina and Dubovsky, Sergei",
    title = "{Exploring the String Axiverse with Precision Black Hole Physics}",
    eprint = "1004.3558",
    archivePrefix = "arXiv",
    primaryClass = "hep-th",
    doi = "10.1103/PhysRevD.83.044026",
    journal = "Phys. Rev. D",
    volume = "83",
    pages = "044026",
    year = "2011"
}

@article{Baryakhtar:2017ngi,
    author = "Baryakhtar, Masha and Lasenby, Robert and Teo, Mae",
    title = "{Black Hole Superradiance Signatures of Ultralight Vectors}",
    eprint = "1704.05081",
    archivePrefix = "arXiv",
    primaryClass = "hep-ph",
    doi = "10.1103/PhysRevD.96.035019",
    journal = "Phys. Rev. D",
    volume = "96",
    number = "3",
    pages = "035019",
    year = "2017"
}

@article{Blas:2020och,
    author = "Blas, D. and Martin Camalich, J. and Oller, J. A.",
    title = "{Scalar resonance in graviton-graviton scattering at high-energies: The graviball}",
    eprint = "2009.07817",
    archivePrefix = "arXiv",
    primaryClass = "hep-th",
    reportNumber = "KCL-2020-54",
    doi = "10.1016/j.physletb.2022.136991",
    journal = "Phys. Lett. B",
    volume = "827",
    pages = "136991",
    year = "2022"
}

@article{Cicoli:2012cy,
    author = "Cicoli, M. and Tasinato, G. and Zavala, I. and Burgess, C. P. and Quevedo, F.",
    title = "{Modulated Reheating and Large Non-Gaussianity in String Cosmology}",
    eprint = "1202.4580",
    archivePrefix = "arXiv",
    primaryClass = "hep-th",
    reportNumber = "DAMTP-2012-14",
    doi = "10.1088/1475-7516/2012/05/039",
    journal = "JCAP",
    volume = "05",
    pages = "039",
    year = "2012"
}

@article{Acharya:2010zx,
    author = "Acharya, Bobby Samir and Bobkov, Konstantin and Kumar, Piyush",
    title = "{An M Theory Solution to the Strong CP Problem and Constraints on the Axiverse}",
    eprint = "1004.5138",
    archivePrefix = "arXiv",
    primaryClass = "hep-th",
    doi = "10.1007/JHEP11(2010)105",
    journal = "JHEP",
    volume = "11",
    pages = "105",
    year = "2010"
}

@article{Sikivie:1983ip,
    author = "Sikivie, P.",
    editor = "Srednicki, M. A.",
    title = "{Experimental Tests of the Invisible Axion}",
    reportNumber = "PRINT-83-0597 (FLORIDA), UF-TP-83-13",
    doi = "10.1103/PhysRevLett.51.1415",
    journal = "Phys. Rev. Lett.",
    volume = "51",
    pages = "1415--1417",
    year = "1983",
    note = "[Erratum: Phys.Rev.Lett. 52, 695 (1984)]"
}

@article{Graham:2015ouw,
    author = "Graham, Peter W. and Irastorza, Igor G. and Lamoreaux, Steven K. and Lindner, Axel and van Bibber, Karl A.",
    title = "{Experimental Searches for the Axion and Axion-Like Particles}",
    eprint = "1602.00039",
    archivePrefix = "arXiv",
    primaryClass = "hep-ex",
    doi = "10.1146/annurev-nucl-102014-022120",
    journal = "Ann. Rev. Nucl. Part. Sci.",
    volume = "65",
    pages = "485--514",
    year = "2015"
}

@article{Irastorza:2018dyq,
    author = "Irastorza, Igor G. and Redondo, Javier",
    title = "{New experimental approaches in the search for axion-like particles}",
    eprint = "1801.08127",
    archivePrefix = "arXiv",
    primaryClass = "hep-ph",
    doi = "10.1016/j.ppnp.2018.05.003",
    journal = "Prog. Part. Nucl. Phys.",
    volume = "102",
    pages = "89--159",
    year = "2018"
}

@article{Sikivie:2006ni,
    author = "Sikivie, Pierre",
    editor = "Kuster, Markus and Raffelt, Georg and Beltran, Berta",
    title = "{Axion Cosmology}",
    eprint = "astro-ph/0610440",
    archivePrefix = "arXiv",
    reportNumber = "UFIFT-HEP-06-16",
    doi = "10.1007/978-3-540-73518-2_2",
    journal = "Lect. Notes Phys.",
    volume = "741",
    pages = "19--50",
    year = "2008"
}

@article{DiLuzio:2020wdo,
    author = "Di Luzio, Luca and Giannotti, Maurizio and Nardi, Enrico and Visinelli, Luca",
    title = "{The landscape of QCD axion models}",
    eprint = "2003.01100",
    archivePrefix = "arXiv",
    primaryClass = "hep-ph",
    reportNumber = "DESY 20-036, DESY-20-036",
    doi = "10.1016/j.physrep.2020.06.002",
    journal = "Phys. Rept.",
    volume = "870",
    pages = "1--117",
    year = "2020"
}

@article{Kim:2008hd,
    author = "Kim, Jihn E. and Carosi, Gianpaolo",
    title = "{Axions and the Strong CP Problem}",
    eprint = "0807.3125",
    archivePrefix = "arXiv",
    primaryClass = "hep-ph",
    doi = "10.1103/RevModPhys.82.557",
    journal = "Rev. Mod. Phys.",
    volume = "82",
    pages = "557--602",
    year = "2010",
    note = "[Erratum: Rev.Mod.Phys. 91, 049902 (2019)]"
}

@article{Turner:1985si,
    author = "Turner, Michael S.",
    title = "{Cosmic and Local Mass Density of Invisible Axions}",
    reportNumber = "FERMILAB-PUB-85-149-A, EFI-85-67-CHICAGO, FERMILAB-PUB-85-093-A",
    doi = "10.1103/PhysRevD.33.889",
    journal = "Phys. Rev. D",
    volume = "33",
    pages = "889--896",
    year = "1986"
}

@article{Visinelli:2009zm,
    author = "Visinelli, Luca and Gondolo, Paolo",
    title = "{Dark Matter Axions Revisited}",
    eprint = "0903.4377",
    archivePrefix = "arXiv",
    primaryClass = "astro-ph.CO",
    doi = "10.1103/PhysRevD.80.035024",
    journal = "Phys. Rev. D",
    volume = "80",
    pages = "035024",
    year = "2009"
}

@article{Grin:2007yg,
    author = "Grin, Daniel and Smith, Tristan L. and Kamionkowski, Marc",
    title = "{Axion constraints in non-standard thermal histories}",
    eprint = "0711.1352",
    archivePrefix = "arXiv",
    primaryClass = "astro-ph",
    doi = "10.1103/PhysRevD.77.085020",
    journal = "Phys. Rev. D",
    volume = "77",
    pages = "085020",
    year = "2008"
}

\end{document}